\documentclass[a4paper,fleqn]{cas-dc}

\usepackage[numbers,sort&compress]{natbib}

\usepackage{algorithm}
\usepackage{algorithmic}
\usepackage[dvipsnames]{xcolor}
\usepackage{todonotes}
\usepackage{multirow}
\usepackage{hhline}

\usepackage{makecell}
\usepackage{pifont}   
\newcommand{\cmark}{\ding{51}} 
\newcommand{\xmark}{\ding{55}} 
\usepackage{adjustbox}
\usepackage{pgfplots}
\usepackage{amsmath,amssymb,amsfonts}
\usepackage{graphicx}
\usepackage{textcomp}
\usepackage{array}
\usepackage{comment}
\usepackage[nolist]{acronym}
\usepackage[normalem]{ulem}
\usepackage{color,soul}
\usepackage{multicol,multirow,colortbl}
\usepackage{subcaption}
\usepackage[dvipsnames]{xcolor}
\usepackage{notoccite}

\def\tsc#1{\csdef{#1}{\textsc{\lowercase{#1}}\xspace}}
\tsc{WGM}

\UseRawInputEncoding

\newdefinition{rmk}{Remark}

\begin{acronym} 
\acro{5G}{Fifth Generation of cellular networks}
\acro{5G NR}{5G new radio}
\acro{6G-V2X}{6G-based vehicle-to-everything}
\acro{AIM}{autonomous intersection management}
\acro{AI}{artificial intelligence}
\acro{AMQP}{advanced message queueing protocol}
\acro{ARQ}{automatic repeat request}
\acro{abstract syntax notation one}{ASN.1}
\acro{BSS}{basic service set}
\acro{C-ITS}{cooperative-intelligent transport systems}
\acro{C-V2X}{cellular-\ac{V2X}}
\acro{CAM}{cooperative awareness message}
\acro{CAV}{connected and autonomous vehicle}
\acro{cdf}{cumulative distribution function}
\acro{CP}{cyclic prefix}
\acro{CPM}{collective perception message}
\acro{CPS}{collective perception service}
\acro{CSMA/CA}{carrier sense multiple access with collision avoidance}
\acro{D2D}{device-to-device}
\acro{DCM}{dual carrier modulation}
\acro{DSRC}{dedicated short range communication}
\acro{EED}{end-to-end delay}
\acro{EGO}{ego-vehicle}
\acro{ETSI}{European telecommunications standards institute}
\acro{FCC}{federal communications commission}
\acro{FCFS}{first-come-first-served}
\acro{FIFO}{first-in-first-out}
\acro{FIFS}{first-in-first-scheduled}
\acro{GLOSA}{green light optimised speed advisory}
\acro{GNSS}{global navigation satellite system}
\acro{KPI}{key performance indicator} 
\acro{LDPC}{low-density parity-check}
\acro{LOS}{line-of-sight}
\acro{LTE}{long term evolution}  
\acro{LTE-V2X}{LTE-based vehicle-to-everything} 
\acro{MAC}{medium access control}
\acro{MEC}{multi-access edge computing}
\acro{MCM}{manoeuvre coordination message}
\acro{MC}{manoeuvre coordination}
\acro{MCS}{manoeuvre coordination service}
\acro{MILP}{mixed-integer linear program}
\acro{ML}{machine learning}
\acro{MOEA}{multi-objective evolutionary algorithm}
\acro{Moveover}{seamless mobility of vehicles over intersections}
\acro{MPC}{model predictive control}
\acro{MSCM}{maneuver sharing and coordinating messages}
\acro{MSCS}{maneuver sharing and coordinating service}
\acro{NLOS}{non-line-of-sight}
\acro{OBU}{on-board unit}
\acro{OCB}{outside of the context of a basic service set}
\acro{OFDM}{orthogonal frequency-division multiplexing}
\acro{PHY}{physical}
\acro{PoC}{proof-of-concept}
\acro{PRB}{physical resource block}
\acro{QAM}{quadrature amplitude modulation}
\acro{QoS}{quality of service}
\acro{RB}{resource block}
\acro{RL}{reinforcement learning}
\acro{RRI}{resource reservation interval}
\acro{RSU}{road-side unit} 
\acro{SAE}{society of automotive engineering}
\acro{SC-FDMA}{single carrier frequency division multiple access}
\acro{SCS}{subcarrier spacing}
\acro{SCI}{sidelink control information}
\acro{SoA}{state-of-the-art}
\acro{SoW}{statement of the work}
\acro{SPS}{semi-persistent scheduling}
\acro{SUMO}{simulation of urban mobility}
\acro{TAI}{time at intersection}
\acro{TRaCI}{traffic control interface}
\acro{TRR}{target road resource}
\acro{UE}{user equipment}
\acro{V2N}{vehicle-to-network}
\acro{V2N2V}{vehicle-to-network-to-vehicle}
\acro{V2I}{vehicle-to-infrastructure} 
\acro{V2V}{vehicle-to-vehicle} 
\acro{V2X}{vehicle-to-everything} 
\acro{veh/s}{vehicles per second}
\acro{VAM}{vulnerable road user awareness message}
\acro{VRU}{vulnerable road user}
\acro{VTL}{virtual traffic light}
\acro{WAVE}{wireless access in vehicular environment}
\end{acronym}
\newcommand{\controlLength}{L}

\newcommand{\speed}[1]{v({{#1}})}
\newcommand{\speedmax}{v_{\text{max}}}
\newcommand{\speedturn}{v_{\text{max-turn}}}

\newcommand{\acceleration}[1]{u({{#1}})}
\newcommand{\accelerationmin}{b_{\text{max}}}
\newcommand{\accelerationmax}{a_{\text{max}}}

\newcommand{\tTAI}{t_\text{INT}}

\newcommand{\tSTART}{t_\text{START}}
\newcommand{\tEND}{t_\text{END}}
\newcommand{\arrivalrate}{\lambda_\text{a}}
\newcommand{\servicetime}{T_\text{s}}
\newcommand{\servicerate}{\mu_\text{s}}
\newcommand{\servicestd}{\sigma_\text{s}}
\newcommand{\utilization}{\rho_\text{u}}
\newcommand{\queuelength}{W_\text{q}}
\newcommand{\queuetime}{T_\text{q}}
\newcommand{\negotiationduration}{T_\text{neg}}
\newcommand{\txduration}{T_\text{x}}
\newcommand{\sojourntime}{T_\text{j}}

\begin{document}
\let\WriteBookmarks\relax
\def\floatpagepagefraction{1}
\def\textpagefraction{.001}

\shorttitle{V2N-Based Algorithm and Communication Protocol for Autonomous Non-Stop Intersections}

\shortauthors{L. Farina, L. M. Amorosa, M. Rapelli, B. M. Masini, C. Casetti, and A. Bazzi}

\title [mode = title]{V2N-Based Algorithm and Communication Protocol for Autonomous Non-Stop Intersections}

\author[1,2]{\textcolor{black}{Lorenzo Farina}}[orcid=0009-0005-9322-8999]
\cormark[1]
\ead{l.farina@unibo.it}
\credit{Conceptualization, Methodology, Software, Writing - original draft}

\author[1,2]{\textcolor{black}{Lorenzo Mario Amorosa}}[orcid=0000-0002-0405-9611]
\ead{lorenzomario.amorosa@unibo.it}
\credit{Methodology, Software, Writing - original draft}

\author[3]{\textcolor{black}{Marco Rapelli}}[orcid=0000-0001-9259-5387]
\ead{marco.rapelli@polito.it}
\credit{Conceptualization, Writing - review \& editing}

\author[4,2]{\textcolor{black}{Barbara Mav\'{i} Masini}}[orcid=0000-0002-1094-1985]
\ead{barbara.masini@cnr.it}
\credit{Conceptualization, Writing - review \& editing}

\author[3]{\textcolor{black}{Claudio Casetti}}[orcid=0000-0002-9507-8526]
\ead{claudio.casetti@polito.it}
\credit{Conceptualization, Writing - review \& editing}

\author[1,2]{\textcolor{black}{Alessandro Bazzi}}[orcid=0000-0003-3500-1997]
\ead{alessandro.bazzi@unibo.it}
\credit{Conceptualization, Methodology, Writing - original draft, Supervision}

\affiliation[1]{organization={DEI, Department of Electrical, Electronic, and Information Engineering ``Guglielmo Marconi''},
    addressline={Universit\`a di Bologna}, 
    city={Bologna},
    postcode={40136},
    country={Italy}
}

\affiliation[2]{organization={WiLab, National Laboratory of Wireless Communications},
    addressline={CNIT}, 
    city={Bologna},
    postcode={40136},
    country={Italy}
}

\affiliation[3]{organization={DAUIN, Department of Control and Computer Engineering},
    addressline={Politecnico di Torino}, 
    city={Torino},
    postcode={10129},
    country={Italy}
}

\affiliation[4]{organization={CNR, National Research Council},
    addressline={IEIIT}, 
    city={Bologna},
    postcode={40136},
    country={Italy}
}

\cortext[1]{Corresponding author}

\begin{abstract}
Intersections are critical areas for road safety and traffic efficiency, accounting for a significant portion of vehicle crashes and fatalities. While connected and autonomous vehicle (CAV) technologies offer a promising solution for autonomous intersection management, many existing proposals either rely on computationally heavy centralized controllers or overlook the practical impairments of real-world communication networks. This paper introduces seamless mobility of vehicles over intersections (Moveover), a novel algorithm comprising a vehicle-to-network (V2N) communication protocol designed to let vehicles cross autonomous intersections without stopping. Moveover delegates trajectory and speed profile selection to individual vehicles, allowing each CAV to optimize them according to its unique kinematic characteristics. Simultaneously, a local intersection controller prevents collisions through deterministic conflict zone reservations. The algorithm is rigorously evaluated under both ideal and non-ideal networking conditions, specifically modeling 4G and 5G communication delays, across multiple layouts including single-lane, multi-lane, and roundabouts. Furthermore, we test Moveover on a real urban map with multiple intersections. Simulation results demonstrate that Moveover significantly outperforms baseline strategies, offering substantial improvements in travel times and reduced pollutant emissions. 
\end{abstract}




\maketitle

\section{Introduction}

According to the fatality analysis reporting system (FARS), 24.8\% of fatal motor vehicle crashes in the United States between 2018 and 2022 occurred at intersections~\cite{fars2022fatality}. In Europe, intersections account for 43\% of injury accidents and 21\% of fatalities~\cite{simon2009intersection}.
These statistics underscore the critical role that intersections play in road safety and highlight the need for targeted measures to mitigate risks in these areas.
Recently, \ac{CAV} technologies have emerged as a promising solution for enhancing road safety and efficiency. Leveraging advancements in \ac{AI}, sensor systems, and \ac{V2X} communications, these technologies have the potential to revolutionize the management of intersections and other critical road segments by improving situational awareness, smoothing traffic flow, and lowering pollutant emissions.

In this context, standardization entities such as the \ac{SAE} and the \ac{ETSI} are playing a pivotal role in shaping the landscape of \ac{V2X} standardization. 
In addition to well-established solutions for cooperative awareness demonstrated through both simulations~\cite{avino2018support} and field tests~\cite{rapelli2024oscar}, recent efforts have focused on developing standardized frameworks to enable maneuver coordination. 
In particular, \ac{SAE} published the \ac{MSCS} in SAE J3186 and J3216~\cite{SAEJ3186,SAEJ3216}, while \ac{ETSI} is working at the \ac{MCS} in the ETSI TR 103 578 and TS 103 561~\cite{etsi103.578,etsi103.561}. These documents define the messages and protocols enabling vehicles to coordinate their actions and include the intersection as one of the main use cases.
It is important to emphasize that these standards do not propose specific traffic management algorithms, instead leaving the selection and implementation of appropriate algorithms to application developers. 

Recognizing the urgency of addressing intersection issues, the research community has mobilized efforts at an international level to explore innovative solutions~\cite{MaksimovskiFestagFacchi2021,GUO2019313,chen2015cooperative,10.1145/3407903,9678327}. However, most proposals in the literature focus primarily on either the communication protocol or the control algorithm for optimizing intersection usage, often oversimplifying or overlooking the other aspect. Differently, in this work we design, implement and validate an algorithm that exploits \ac{V2N} communications to enable vehicles crossing the intersections without stopping, unless the road network is congested. 
The contribution of this work can be summarized as follows:
\begin{itemize}
  \item We propose a novel algorithm, called \ac{Moveover}, 
  to define the movements of vehicles at signal-free intersections which 1) explicitly accounts for turning maneuvers at crossroads, 2) allows each CAV to independently select a motion pattern that respects its own kinematic and dynamic characteristics, and 3) minimizes the information exchange and avoids heavy online optimization tools (e.g., \ac{MILP} solvers) and learning-based fine-tuning (e.g., \ac{RL}), allowing to scale to large and dense intersections. 
  The proposed algorithm is fully explainable and deterministic, meaning that every scheduling decision and rejection can be traced to explicit reservation checks rather than opaque black-box models.
  \item We specify a practical and lightweight \ac{V2N} communication protocol between vehicles and a local intersection controller that implements the algorithm.
  \item We evaluate the algorithm under non-ideal networking conditions and on multiple intersection layouts (including roundabouts and multi-lane crossroads), showing improved travel time and reduced emissions in simulation compared to baseline strategies.
  \item We provide open source code for the algorithm, the controller, and the simulation scenarios to facilitate reproducibility and further development.\footnote{The code is available at \href{https://github.com/V2Xgithub/Moveover}{https://github.com/V2Xgithub/Moveover}.}
\end{itemize}

This article substantially extends the preliminary concepts introduced in~\cite{10575944}. While the earlier work established the core algorithmic foundation, the present study introduces several key advancements. Specifically, we have comprehensively revised the terminology and simplified the underlying assumptions for greater clarity. Furthermore, we formalize the V2N message set to align with SAE guidelines~\cite{SAEJ3186}, explicitly evaluate the algorithm's robustness under non-ideal communication settings, and rigorously validate its performance across multiple, heterogeneous intersection layouts.

The remainder of this article is organized as follows. Section~\ref{sec:SoA} surveys related work on intersection management, reservation and trajectory-planning approaches and highlights gaps that motivate our proposal. Section~\ref{sec:Intersection_modelling} formalizes the intersection model and introduces trajectory negotiation. Section~\ref{sec:commprotocol} specifies the V2N communication protocol. Section~\ref{sec:commImpact} analyzes the impact of non-ideal communications on the system. Section~\ref{sec:simulation} presents the simulation setup and discusses results obtained on multiple intersection layouts. 
Section~\ref{sec:conclusions} draws the conclusions of the paper.
Finally, Appendix~\ref{sec:zonedesign} and Appendix~\ref{sec:ETSImessage} provide details on the proposed algorithm and communication protocol.

\section{Related works}\label{sec:SoA}

Traffic management at intersections has been widely researched in recent years. Table \ref{tab:sota} summarizes the various approaches proposed in the literature, which we discuss in detail below.
Early studies focused on traffic-light-controlled intersections, exploring two main strategies to improve traffic efficiency. The first aims to optimize traffic-light cycles by minimizing performance indicators such as delays, number of stops, fuel consumption, and exhaust emissions~\cite{li2004signal,stevanovic2009optimizing}. The second aims to optimize vehicle trajectories by providing speed advisories computed through \ac{V2I} communication with a \ac{RSU} connected to the traffic light, thereby smoothing traffic flow and avoiding stop-and-go manoeuvres~\cite{5309519,5982524}. However, these solutions are not applicable in the absence of infrastructure, whose deployment at all unregulated intersections is clearly unfeasible due to high costs.

More recently, it has been shown that transitioning from traffic light controlled intersections to non-signalized ones, which are frequently referred to in the literature as autonomous intersections, can provide substantial benefits in terms of intersection throughput and travel time \cite{tachet2016revisiting}. This shift, along with advancements in \ac{AI}, CAV technologies, and V2X communications, has significantly prompted research addressing \ac{AIM}. 
Among relevant surveys on maneuver coordination, \cite{chen2015cooperative,10.1145/3407903,9678327}  specifically focus on traffic management at intersections. In particular, \cite{9678327} presents use cases for \ac{AIM}, discussing the possible coordination architectures, scheduling policies, intersection modeling and major goals, identified in safety, efficiency, infotainment and environment. As the authors highlight, the majority of proposals tackles safety and efficiency aspects, resorting to different scheduling techniques. Among these, a reservation-based intersection control mechanism is proposed in \cite{dresner2004multiagent}, in which the intersection is subdivided into slots that can be reserved by vehicles querying an intersection manager in a \ac{FCFS} fashion. The same scheduling policy is employed and further developed in \cite{wang2021digital}, which defines an enhanced \ac{FIFO} slot reservation algorithm that relaxes the strict \ac{FIFO} structure and introduces re-scheduling. 

Virtual traffic light algorithms are discussed in~\cite{BAZZI201642} and \cite{JAME2022167}. In the first, a vehicle approaching the intersection can be granted the priority to cross it, and, while performing the crossing manoeuvre, it forwards the priority to another vehicle based on the proposed algorithm; in the second, a procedure is defined to let vehicles self-organize into clusters and establish a crossing order. In \cite{8370718}, an intersection controller periodically optimizes vehicles arrival times at the junction by solving a \ac{MILP}, and each vehicle autonomously plans its trajectory to comply with the assigned time. Authors in \cite{lin2019graph} extract a passing order from a directed graph representing vehicles approaching the intersection and their possible collisions, while in \cite{a15090326} \ac{RL} is applied to learn a scheduling policy that assigns crossing priorities to vehicles; a combination of both approaches is presented in \cite{klimke2022enhanced}. While effective, \ac{MILP} and \ac{ML} solutions often rely on a central controller performing online optimization or learning-based tuning. This not only poses significant scalability challenges in high-density scenarios, but also introduces a degree of opacity, as the underlying logic of black-box models can be difficult to trace.

    	\begin{table*}[!t]
	\caption{Literature overview.}\label{tab:sota}
		\centering 
		\small 
		\thispagestyle{empty}
    \makebox[\textwidth][c]{
	\begin{tabular}{c|c|c|c|c|c|c}
\hline \hline
\textbf{Reference} &\makecell{\textbf{Computational} \\ \textbf{scalability}} & \makecell{\textbf{Vehicle profile} \\ \textbf{control}} & \makecell{\textbf{Emissions} \\ \textbf{reduction}} & \makecell{\textbf{Non-ideal} \\ \textbf{comms.}} & \makecell{\textbf{Comm. impact}\\ \textbf{assessment}} & \textbf{Covered scenarios} \\ \hline \hline
\makecell{Li \textit{et al.} \cite{li2004signal}} & \cmark & \xmark & \cmark & \xmark & \xmark & Three-way multiple-lanes \\ \hline
\makecell{Stevanovic \textit{et al.} \cite{stevanovic2009optimizing}} & \cmark & \xmark & \cmark & \xmark & \xmark & Multiple intersections network \\ \hline
\makecell{Mandava \textit{et al.} \cite{5309519}} & \cmark & \xmark & \cmark & \xmark & \xmark & Multiple intersections network \\ \hline
\makecell{Katsaros \textit{et al.} \cite{5982524}} & \cmark & \xmark & \cmark & \xmark & \xmark & Multiple intersections network \\ \hline
\makecell{Dresner and Stone \cite{dresner2004multiagent}} & \cmark & \makecell{\xmark} & \xmark & \xmark & \xmark & \makecell{Four-way single-lane \\ Four-way multiple-lanes}\\ \hline
\makecell{Wang \textit{et al.} \cite{wang2021digital}} & \cmark & \xmark & \xmark & \xmark & \xmark & Four-way multiple-lanes \\ \hline
\makecell{Bazzi \textit{et al.} \cite{BAZZI201642}} & \cmark & \xmark & \xmark & \cmark & \cmark & Four-way single-lane \\ \hline
\makecell{Jame \textit{et al.} \cite{JAME2022167}} & \cmark  & \xmark & \cmark & \cmark & \xmark & \makecell{Four-way single-lane \\ Three-way single-lane \\ Multiple intersections network} \\ \hline
\makecell{Fayazi and Vahidi \cite{8370718}} & \xmark & \xmark & \xmark & \xmark & \xmark & Four-way single-lane \\ \hline
\makecell{Lin \textit{et al.} \cite{lin2019graph}} & \xmark & \xmark & \xmark & \xmark & \xmark & Four-way single-lane \\ \hline
\makecell{Karthikeyan \textit{et al.} \cite{a15090326}} & \cmark & \xmark & \xmark & \xmark & \xmark & Four-way multiple-lanes \\ \hline
\makecell{Klimke \textit{et al.} \cite{klimke2022enhanced}} & \xmark & \xmark & \xmark & \xmark & \xmark & \makecell{Four-way single-lane \\ Four-way multiple-lanes \\ Three-way single-lane} \\ \hline
\makecell{Zhang and \\ Cassandras \cite{zhang2019decentralized}} & \cmark & \cmark & \cmark & \xmark & \xmark & Four-way single-lane \\ \hline
\makecell{Chen \textit{et al.} \cite{chen2020optimal}} & \cmark & \cmark & \cmark & \xmark & \xmark & Four-way single-lane \\ \hline
\makecell{Hult \textit{et al.} \cite{hult2018energy}} & \xmark & \cmark & \cmark & \xmark & \xmark & Four-way single-lane \\ \hline
\makecell{Martin-Gasulla and \\ Elefteriadou \cite{martin2021traffic}} & \xmark & \cmark & \xmark & \xmark & \xmark & Roundabout \\ \hline
\makecell{Ripon \textit{et al.} \cite{ripon2017multi}} & \xmark & \cmark & \cmark & \xmark & \xmark & Four-way multiple-lanes \\ \hline
\makecell{Zheng \textit{et al.} \cite{zheng2017delay}} & \cmark & \xmark & \xmark & \cmark & \xmark & Four-way single-lane \\ \hline
\makecell{Lu \textit{et al.} \cite{lu2021optimization}} & \xmark & \cmark & \xmark & \cmark & \xmark & Four-way multiple-lanes \\ \hline
\makecell{Chamideh \textit{et al.} \cite{chamideh2022safe}} & \cmark & \cmark & \cmark & \cmark & \xmark & Four-way single-lane \\ \hline
\makecell{This \\ work} & \cmark & \cmark & \cmark & \cmark & \cmark & \makecell{Four-way single-lane \\ Four-way multiple-lanes \\ Three-way single-lane \\ Roundabout \\ Multiple intersections network} \\ \hline
\end{tabular}}
	\end{table*}

Beyond these architectural and transparency issues, a common limitation of the previously discussed solutions is that they primarily focus on defining scheduling policies to ensure safety and improve traffic efficiency, but they do not directly intervene in the control inputs or \emph{mobility profiles}\footnote{For conciseness, we use the term mobility profile to denote a vehicle's trajectory together with its speed profile.} of the vehicles. This, however, could yield substantial benefits in terms of both pollutant emissions and energy consumption reduction, which is another crucial aspect when discussing \ac{AIM}. Although it is clear that smoothing traffic flow and reducing delays already have a positive impact in this regard, a further step is to define a control framework that computes vehicle trajectories to minimize their environmental impact. To this end, it is essential that vehicles maintain a constant speed as much as possible and avoid stop-and-go maneuvers~\cite{TRL384}.

Several works in the literature discuss how to derive proper control inputs to handle vehicle trajectories. In \cite{zhang2019decentralized}, a central coordinator computes arrival times at the intersection of vehicles inside a control zone, and then each vehicle computes its own acceleration profile by minimizing a cost function that is related to the energy consumption. In \cite{chen2020optimal} instead, the profile is directly computed by a central controller, performing a joint minimization of travel time and energy consumption. A similar approach is considered in \cite{hult2018energy}, but \ac{MPC} is performed to update the profile periodically. In \cite{martin2021traffic}, the profile periodic recalculation is based solely on time constraints related to collisions. Authors in \cite{ripon2017multi} employ a \ac{MOEA} to calculate vehicle speed profiles that maximize throughput and minimize mean evacuation time, total loss of kinetic energy, and number of collisions. Unlike the previously mentioned \ac{ML} approaches, the majority of these works avoid black-box models in favor of fully explainable deterministic frameworks. Nevertheless, scalability remains a concern for those relying on central controllers that employ \ac{MPC}, which may struggle to handle the computational demand of high-traffic scenarios.

Apart from the control framework, there is still a significant shortcoming that characterizes the majority of works addressing \ac{AIM}. The performance of the proposed solutions is typically evaluated assuming that communications among vehicles and with external infrastructures take place without any impairment. Nevertheless, this assumption becomes unrealistic when considering a practical implementation of the proposal. In particular, the same levels of safety and efficiency cannot be guaranteed when delays and packet losses occur, and the communication network may not be able to handle the wireless traffic generated when high vehicle densities are present in the scenario, especially when large amounts of data need to be exchanged between vehicles and a central controller. According to \cite{wang2024assessing}, \ac{AIM} systems fail to outperform common traffic lights when communication delays are higher than 247 ms.

Very few studies on \ac{AIM} assess the impact of real communications on the effectiveness of the proposed solutions. In \cite{zheng2017delay}, a communication protocol involving vehicles and a central controller is defined to tolerate delays, but the outcome of the procedure is limited to determining the intersection crossing order, with no focus on vehicle profiles. The latter is instead encompassed in \cite{lu2021optimization}, where a \ac{MILP} is detailed to minimize travel times under imperfect \ac{V2I} communications. As previously discussed, however, such centralized \ac{MILP}-based frameworks can become a computational bottleneck as vehicle density increases. Authors in \cite{chamideh2022safe} employ \ac{MPC} to minimize the energy consumption. A safe speed for all vehicles is periodically computed and communicated by a central controller, but vehicles can also monitor their surroundings through sensors and make autonomous decisions when not receiving instructions. Such approach, called hierarchical \ac{MPC}, divides the optimization control problem in two layers, achieving a reduction of the computational complexity.

Although it is true that the aforementioned works define clear communication protocols and conduct some simulations with non-ideal communications, they do not carry out a thorough evaluation of the impact of the communication protocol on the network.
It is also important to highlight that most proposals in the literature validate their algorithms on a single type of intersection, without investigating whether they could be generalized to different types of intersections or applied to multiple intersections scenarios.

\section{Moveover: overview and building blocks}\label{sec:Intersection_modelling}

The proposed \ac{Moveover} algorithm aims to be general, without setting any specific requirement on the intersection type and structure, and allowing a variable number of ways and lanes. We model an intersection as a finite set of roads connecting at a common area. Each road may contain one or more \emph{lanes}. We denote by $\mathcal{L}$ the set of \emph{incoming lanes} (lanes that enter the intersection) and by $\mathcal{O}$ the set of \emph{outgoing lanes}. A \emph{path} is an ordered sequence of road segments connecting an incoming lane $\ell_{\text{in}}\in\mathcal{L}$ to an outgoing lane $\ell_{\text{out}}\in\mathcal{O}$.
The set of possible paths is finite and known, based on the road map and turn permissions. Similarly, also the areas where collisions may occur vary across intersections, but they are always finite and known.
Hereafter, the key aspects of \ac{Moveover} are discussed. The main building blocks of \ac{Moveover} are defined and summarized in Table~\ref{tab:defs}.

    	\begin{table}[t]
	\caption{Building blocks of the proposed algorithm. 
    }\label{tab:defs}
		\centering 
	\begin{tabular}{p{2.8cm}|p{5.2cm}}
\hline \hline
\makecell{\textbf{Term}} & \makecell{\textbf{Meaning}} \\ \hline \hline
Conflict zone & Portion of road that can be used by a single CAV at a time \\  \hline
Controller & Software entity that controls the intersection and accepts or modifies the mobility profiles proposed by the CAVs\\  \hline
EGO & CAV that is negotiating its mobility profile with the controller \\  \hline
\hbox{Minimum negotiation} distance & Minimum distance between the end of the negotiation zone and the entrance of the intersection \\ \hline
Mobility profile & Trajectory and speed profile of a CAV \\  \hline
Negotiation length & Distance between the starting of the negotiation zone and its end \\  \hline
Negotiation zone & Area in a lane, before the intersection, within which the mobility profile is agreed between the CAVs and the controller \\  \hline

\hbox{Scheduling table} & Table stored by the controller with the reservations of the conflict zones \\ 
\hline
\hline
\end{tabular}
	\end{table}

\begin{figure*}[t]
\centering
\subfloat[]{\includegraphics[width=0.48\columnwidth]{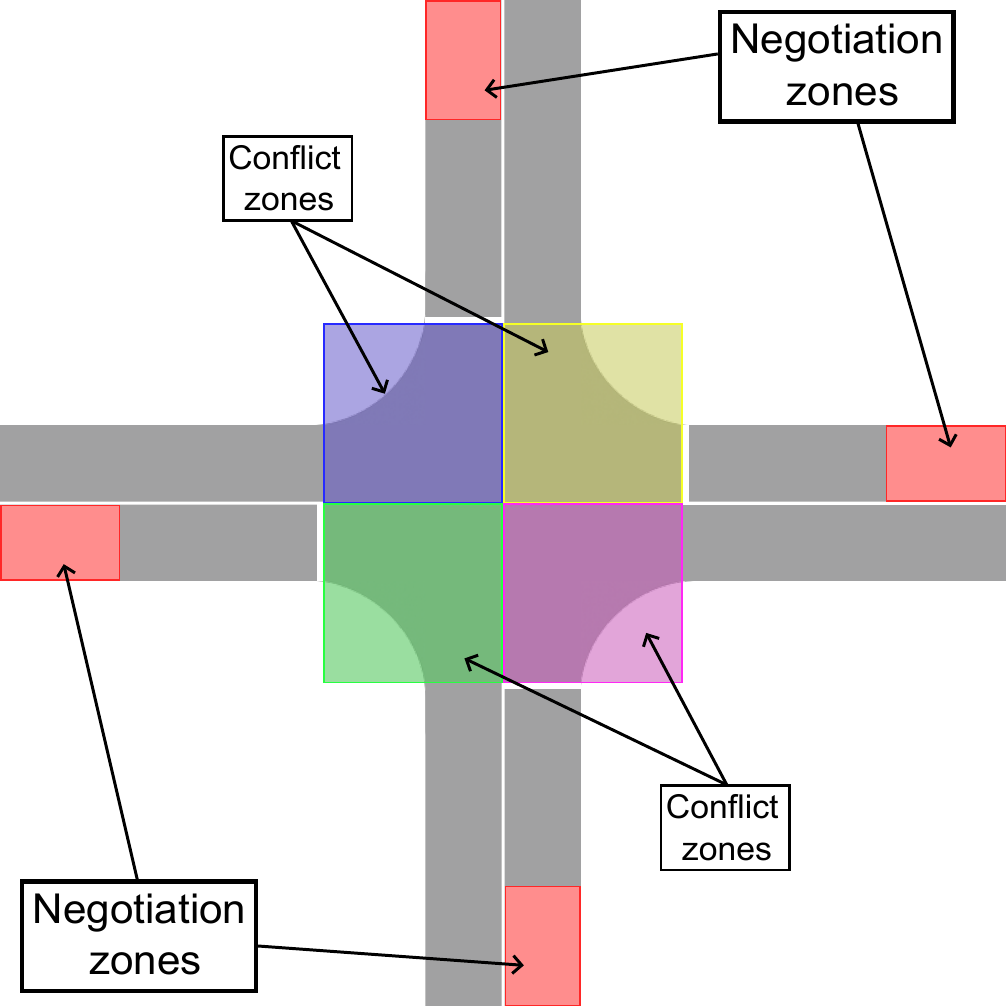}
\label{fig:model_a}}
\hfill
\subfloat[]{\includegraphics[width=0.48\columnwidth]{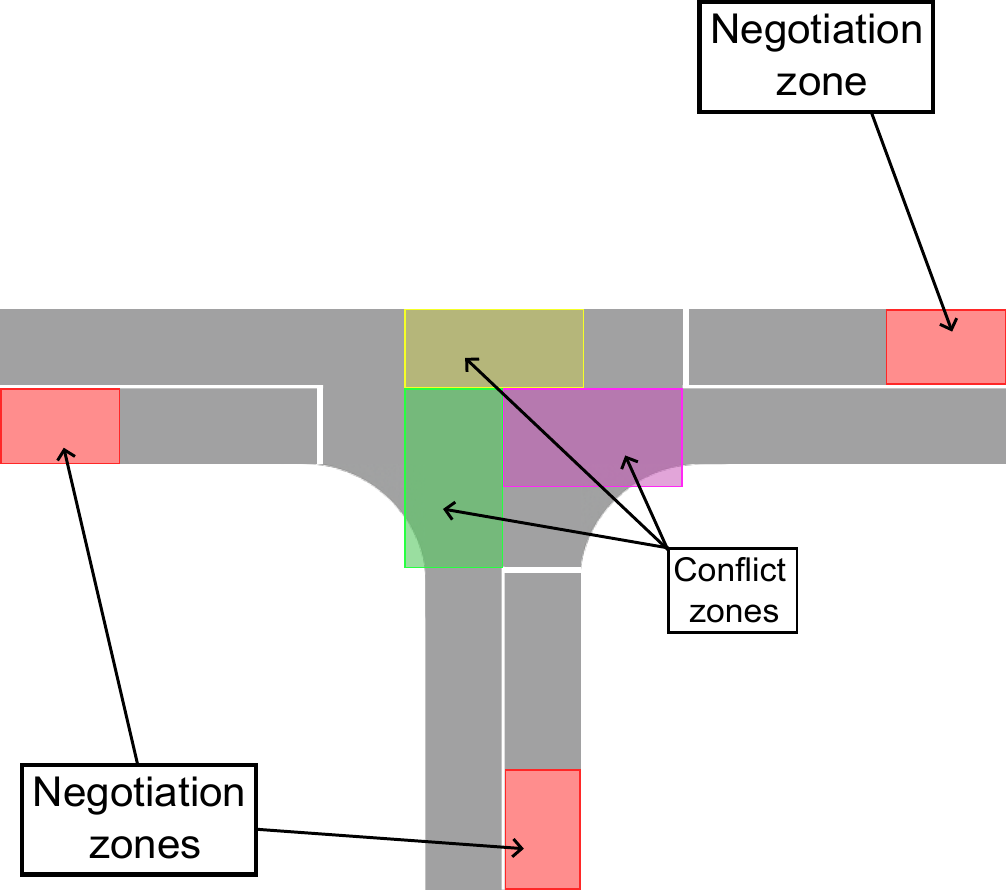}
\label{fig:model_b}}
\hfill
\subfloat[]{\includegraphics[width=0.48\columnwidth]{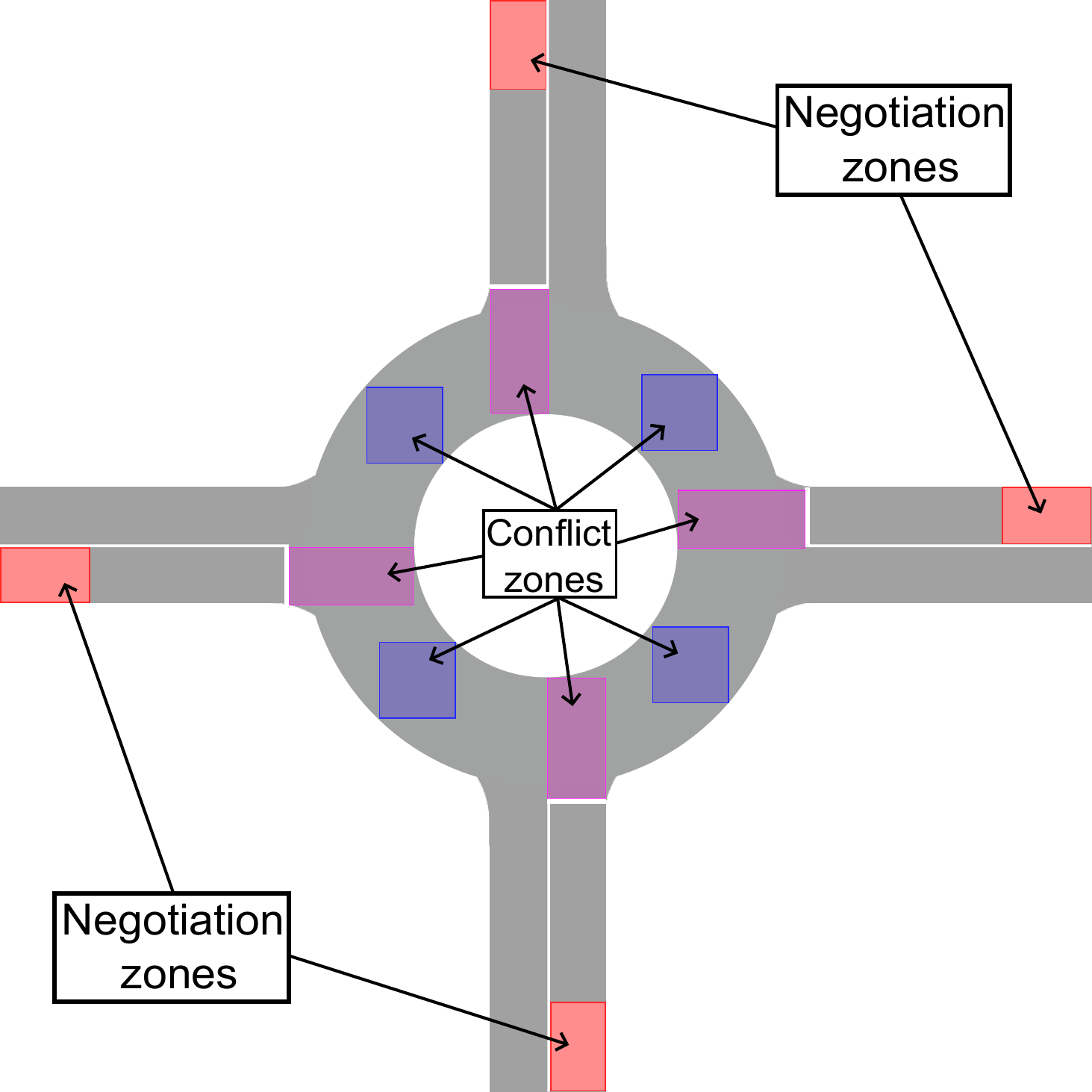}
\label{fig:model_c}}
\hfill
\subfloat[]{\includegraphics[width=0.48\columnwidth]{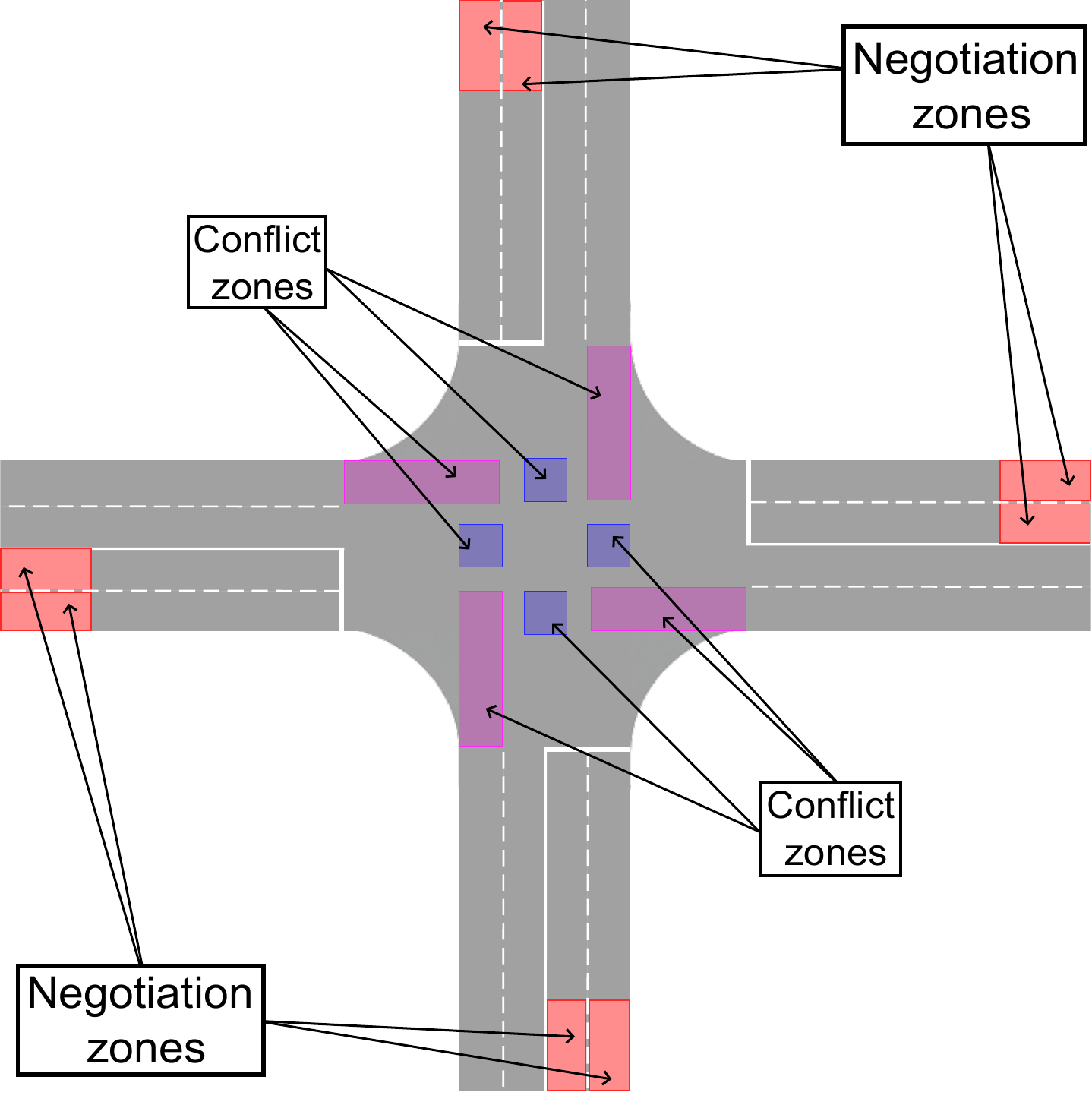}
\label{fig:model_d}}
\caption{Examples of considered intersections with negotiation and conflict zones. (a) Four-way single-lane intersection. (b) Three-way single-lane intersection. (c) Roundabout. (d) Four-way two-lane intersection.} 
\label{fig:model}
\end{figure*}

\subsection{Core principles of the algorithm }
 
The intersection is managed by a \textit{controller}; all \acp{CAV} are connected to it via \ac{V2N} communication. This can be implemented in several ways, such as through \acp{RSU} equipped with either IEEE~802.11p-based systems or cellular sidelink wireless technologies. In this work, we assume the availability of cellular coverage with both uplink and downlink connectivity, and that the controller is implemented on a \ac{MEC} server to minimize the latency between the CAVs and the controller. 

\ac{Moveover} runs in part on CAVs and in part at the controller. 
The purpose of this algorithm is to determine, for each CAV, a mobility profile that enables it to traverse the intersection without collisions and without stopping, except in cases of road congestion. The mobility profile is determined based on: 1) the characteristics of each CAV, known only by the CAV itself; and 2) the mobility profiles of the other CAVs, known only by the controller. 

Mobility profiles are computed sequentially for one vehicle at a time: at any time, the vehicle actively negotiating its profile with the controller is referred to as \textit{EGO}. Once the mobility profile is assigned to a CAV, it remains fixed unless the system switches to the exceptional status of backup mode, as detailed later. 
A key strength of this approach is its minimal impact on channel load, which makes it bandwidth-efficient and scalable to large intersections and complex urban environments.

\begin{rmk}
The intersection is managed by the 
controller, but
it is in charge of each EGO to calculate its own mobility profile (not considering other CAVs). 
This approach differs from many methods in the literature, where the controller is responsible for computing the mobility trajectories of all vehicles. By delegating trajectory optimization to each vehicle, our method simplifies the implementation and better accounts for heterogeneous vehicle characteristics. For instance, the dynamics of a van can differ considerably from those of a sports car or a city car, and significant differences can exist even among different models of similar vehicles. 
\end{rmk}

\subsection{Negotiation and conflict zones}

\ac{Moveover} relies on the following modeling of the intersection. Each incoming lane is associated with a \textit{negotiation zone}, and the intersection itself contains several \textit{conflict zones}: 
\begin{itemize}
    \item The \textit{negotiation zone} is the area where the EGO and the controller coordinate to establish the EGO mobility profile. Each incoming lane has its own negotiation zone, and the mobility profile must be finalized before the EGO leaves this area. If the process is not completed in time, the system switches to backup mode, as discussed in Section \ref{sec:backup_mode_section}.
    \item A \textit{conflict zone} is a road segment that can be occupied by only one CAV at a time to prevent collisions. Safety is ensured by enforcing strictly non-overlapping temporal reservations for each of these zones.
\end{itemize}
When a CAV enters a negotiation zone, it begins negotiating its mobility profile with the controller. The CAV becomes the EGO once all previously entered CAVs have finalized their mobility profiles. All CAVs are therefore served using a \ac{FIFS} approach.
Within each negotiation zone, a vehicle must hold a predetermined constant speed and the negotiated mobility profile is applied when exiting from that zone. 

The size of the negotiation zones, as well as the number and shape of the conflict zones, depend on the specific intersection and are subject to design choices, as further elaborated in Appendix~\ref{sec:zonedesign}.
As examples of intersections, those considered in our work are shown with negotiation and conflict zones  in Fig.~\ref{fig:model}.

\subsection{Coordinating messages}\label{sec:messages}
\ac{Moveover} negotiation relies on two types of messages:
\begin{itemize}
    \item A \textit{message sent by the EGO to the controller}, containing the proposed mobility profile; the mobility profile represented as a sequence of positions, each annotated with the corresponding time instant at which it is reached. These positions must span the entire path from the negotiation zone to the exit of the last traversed conflict zone.
    \item A \textit{message sent by the controller to the EGO}, containing the possible time reservations of each traversed conflict zone; specifically, for each traversed conflict zone, the minimum instant at which the EGO may enter and the maximum instant by which it must exit.
\end{itemize}
While standardization bodies such as SAE and ETSI are still defining \acp{MCM} for maneuver coordination \cite{SAEJ3186,SAEJ3216,etsi103.578,etsi103.561}, Appendix~\ref{sec:ETSImessage} outlines a potential ETSI-compliant implementation. This approach utilizes the \ac{TRR} concept to represent both the mobility profiles and the conflict zones.

    	\begin{table*}[t]
	\caption{Example of a scheduling table, where three vehicles (V1, V2, and V3) are already scheduled and the controller needs
to schedule V4. 
}\label{tab:schedule}
		\centering 
		\scriptsize
		\thispagestyle{empty}
        \renewcommand{\arraystretch}{1.8} 
    \makebox[\textwidth][c]{
	\begin{tabular}{c|c|c|c|c|c|c|c|c|c|c|c}
\hline \hline
\textbf{CAV ID} & \textbf{Trajectory} & \makecell{\textbf{Entry} \\ \textbf{road}} & \makecell{\textbf{Exit} \\ \textbf{road}} & \makecell{\textbf{Entering} \\ \textbf{ 1}} & \makecell{\textbf{Exiting} \\ \textbf{1}} & \makecell{\textbf{Entering} \\ \textbf{2}} & \makecell{\textbf{Exiting} \\ \textbf{2}} & \makecell{\textbf{Entering} \\ \textbf{3}} & \makecell{\textbf{Exiting} \\ \textbf{3}} & \makecell{\textbf{Entering} \\ \textbf{4}} & \makecell{\textbf{Exiting} \\ \textbf{4}} \\ \hline \hline
Scheduled: V1 & (x$_1$,y$_1$,t$_1$),...,(x$_N$,y$_N$,t$_N$) & West & East & 152 & 153.4 & 153 & 153.9 & & & & \\ \hline
Scheduled: V2 & (x$_1$,y$_1$,t$_1$),...,(x$_N$,y$_N$,t$_N$) & East & West & & & & & 153 & 153.9 & 153.5 & 154.4 \\ \hline
Scheduled: V3 & (x$_1$,y$_1$,t$_1$),...,(x$_N$,y$_N$,t$_N$) & South & North & & & 154.5 & 155.4 & 155 & 155.9 & & \\ \hline
New: V4 & (x$_1$,y$_1$,t$_1$),...,(x$_N$,y$_N$,t$_N$) & East & North & & & & & \makecell{Proposal:\\ \underline{153.5}} & \underline{154.5} & & \\ \hline
\end{tabular}}
	\end{table*}

\subsection{The scheduling table of the controller}\label{sec:schedulingTable}

The controller needs to record the negotiated mobility profiles until the corresponding \acp{CAV} exit from the intersection. An effective way for the controller to store this information and use it during the negotiation of the mobility profiles of new vehicles entering the intersection is the use of a table which we refer to as \textit{scheduling table}.

The scheduling table includes: (i) a column indicating the CAV ID; (ii) a column specifying the trajectory as a sequence of positions and time instants; (iii) one column indicating the entry road to the intersection and one for the exit road; and (iv) for each conflict zone, one column recording the entry time and one column recording the exit time. The number of rows of the table corresponds to the number of CAVs with an assigned mobility profile that have not yet left their last conflict zone. An example is provided in Table~\ref{tab:schedule}.

The entry and exit lane columns allow the controller to immediately identify potentially conflicting CAVs before and after the intersection. Similarly, the columns related to the conflict zones make collision avoidance straightforward: the only requirement is that no entry time into a conflict zone overlaps with the interval between another CAV's entry and exit in the same zone. Additional details on how the controller processes this information are provided in Section~\ref{sec:controllervalidation}.

\section{Moveover: protocol details}\label{sec:commprotocol}

The communication protocol of \ac{Moveover} includes a proposal from the EGO and a response from the controller, possibly followed by additional exchanges until the mobility profile is agreed or the backup mode is entered. 
This procedure is hereafter detailed.

\subsection{Proposal by the EGO} When the EGO enters the negotiation zone, it initially proposes a mobility profile without taking into account the possible presence of other CAVs. Since the EGO knows its own capabilities and intended direction, it has all the necessary information to compute an accurate profile. In particular, the EGO calculates the mobility profile such that:
\begin{itemize}
    \item $\speed{t} \leq \speedmax$ $\forall t \in [\tSTART, \tEND]$, where $\speedmax$ is the road speed limit, $\tSTART$ is the initial instant of the proposal, and $\tEND$ is its last instant;
    \item If the EGO performs a right or left turn, $\speed{t}\leq \speedturn$ $\forall t \in [\tTAI, \tEND]$, where $\tTAI$ is the arrival time at the intersection and $\speedturn<\speedmax$ denotes the maximum safe and comfortable turning speed, specific to each CAV;
    \item The acceleration satisfies $\accelerationmin\leq \acceleration{t}\leq \accelerationmax$ $\forall t \in [\tSTART, \tEND]$, where $\accelerationmin < 0$ and $\accelerationmax > 0$ are the maximum allowed deceleration and acceleration of the EGO, respectively. These parameters depend on the characteristics of the specific CAV, with $\accelerationmin$ representing a comfort-based deceleration threshold rather than the physical braking limit of the vehicle. 
\end{itemize}
The calculated mobility profile is proposed to the controller using the message described in Section~\ref{sec:messages}.

\begin{figure*}[t!]
\centering
\begin{subfigure}[b]{0.30\textwidth}
    \includegraphics[width=\textwidth]{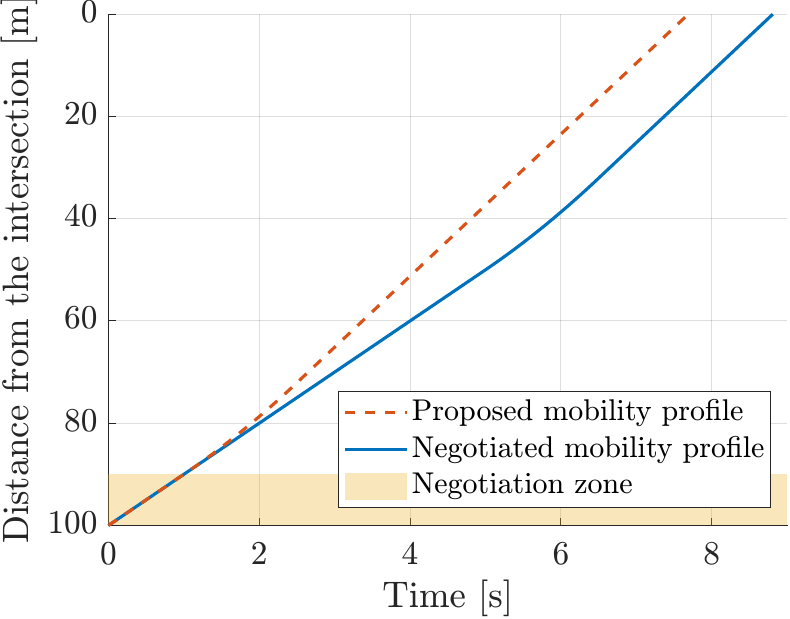}
    \caption{Example position profile}
    \label{fig:distanceProfile}
\end{subfigure}
\hfill
\begin{subfigure}[b]{0.30\textwidth}
    \includegraphics[width=\textwidth]{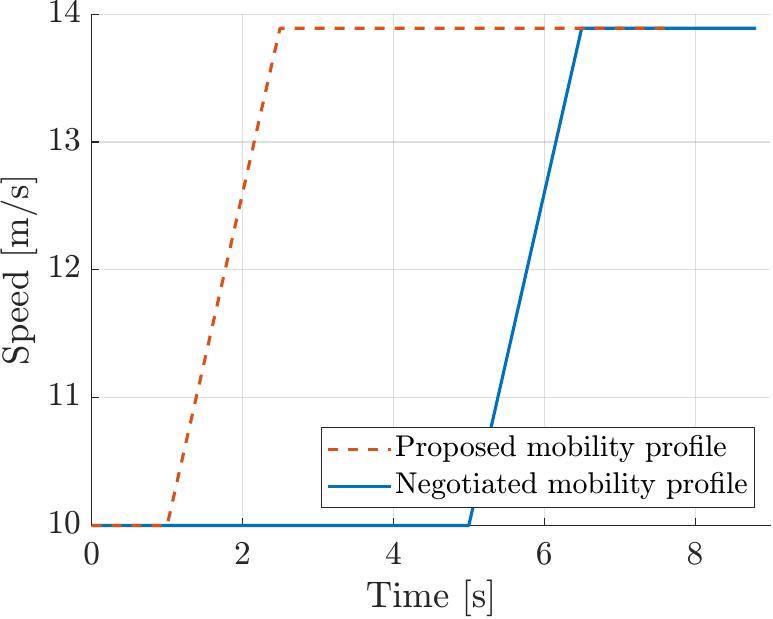}
    \caption{Example speed profile}
    \label{fig:speedProfile}
\end{subfigure}
\hfill
\begin{subfigure}[b]{0.30\textwidth}
    \includegraphics[width=\textwidth]{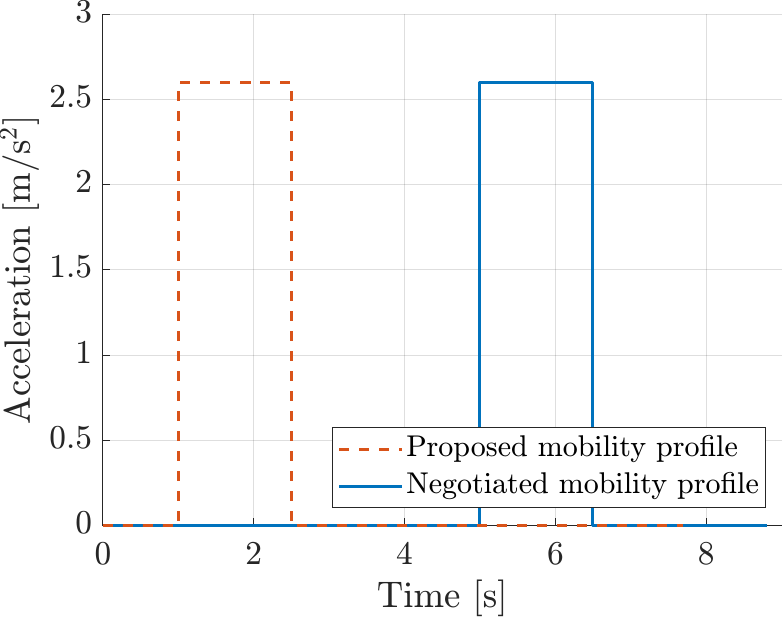}
    \caption{Example acceleration profile}
    \label{fig:accelProfile}
\end{subfigure}
\caption{An example of mobility profile planning performed by the EGO with the setting of the motion profiles, assuming a negotiation length of 10 meters.}
\label{fig:exampleProfile}
\end{figure*}

\subsection{Validation by the controller}\label{sec:controllervalidation} Once the controller receives the proposal, it verifies whether the mobility profile conflicts with those of the CAVs that have already finalized their profiles prior to the EGO, which are stored in the scheduling table.

The controller's validation procedure consists of the following three steps:

    \subsubsection{Conflicts before the intersection} The controller uses the third column of the scheduling table to identify the last CAV that entered the same lane as the EGO, referred to as the \textit{ahead entering CAV}. If such a vehicle exists, the controller checks, using the second column, that the EGO and the ahead entering CAV remain separated by at least a minimum safety distance at all times. If this condition is not satisfied, the controller estimates the delay that the EGO should introduce before it can traverse the conflict zones.

    \subsubsection{Conflicts within the intersection} The controller examines the columns corresponding to each conflict zone to ensure that the EGO's trajectory does not intersect any zone currently reserved by another CAV. If a conflict is detected, the controller postpones the reservation of the conflict zones to the earliest time intervals in which they are free of use by other CAVs.

    \subsubsection{Conflicts after the intersection} The controller uses the fourth column of the scheduling table to identify the last CAV that will exit the same lane before the EGO, referred to as the \textit{ahead exiting CAV}. If such a vehicle exists, the controller verifies that, upon exiting the conflict zone it traverses, the EGO's speed and its distance from the ahead exiting CAV are compatible with the latter's speed. If this condition is not satisfied, the controller estimates the delay that the EGO should introduce before it can traverse the conflict zones, and then repeats the conflict checks before and within the intersection. 
    Note that other CAVs that have been already scheduled may, in principle, exit the same lane after the EGO; in such cases, the EGO necessarily leaves the intersection at a higher speed than those already scheduled, and no further checks are required.

\subsection{Response from the controller}

The response message includes time intervals for each conflict zone, as described in Section~\ref{sec:messages}. The acceptance or rejection of the EGO's proposal is implicit in this response: if the reserved time intervals for the conflict zones are compatible with the proposed profile, the EGO infers that the proposal has been accepted; otherwise, it deduces that the reservations are not compatible and uses this information to calculate a new profile. 

Upon accepting the proposed profile, the controller determines a time interval for each conflict zone, defined by the earliest entry time and the latest exit time. To accommodate potential inaccuracies in the mobility profile, this interval can be widened (lowering the entry time and raising the exit time) to create a safety margin. These bounds are recorded in the scheduling table, as detailed in Section~\ref{sec:schedulingTable}. 

If the proposal is rejected, the controller specifies the available time intervals for each conflict zone, factored against currently scheduled profiles. If no subsequent profiles occupy a zone, the upper time bound is set to infinity. Since the negotiation remains active, these tentative values are not recorded in the scheduling table.

\subsection{Additional exchanges} 
Upon rejection, the EGO attempts to compute a new mobility profile that adheres to the availability windows specified for each conflict zone. If strict adherence is infeasible (for instance, if the required velocity falls below the EGO's minimum speed) the EGO submits a fallback profile satisfying only the entry time of the first conflict zone. The controller identifies this persistent incompatibility and subsequently postpones the scheduling attempt to a later time interval. 

In principle, this negotiation process could iterate multiple times. However, since each response from the controller can only postpone the entry time into the first conflict zone, the number of exchanges remains small. 
Fig.~\ref{fig:exampleProfile} shows an example of mobility profile planning performed by the EGO. The initial proposal, which foresees accelerating to maximum speed immediately outside the negotiation zone, is rejected by the controller. Consequently, the EGO computes a new profile that delays the acceleration, which is eventually accepted by the controller.

\begin{figure*}[t]
\centering
\includegraphics[width=1.4\columnwidth]{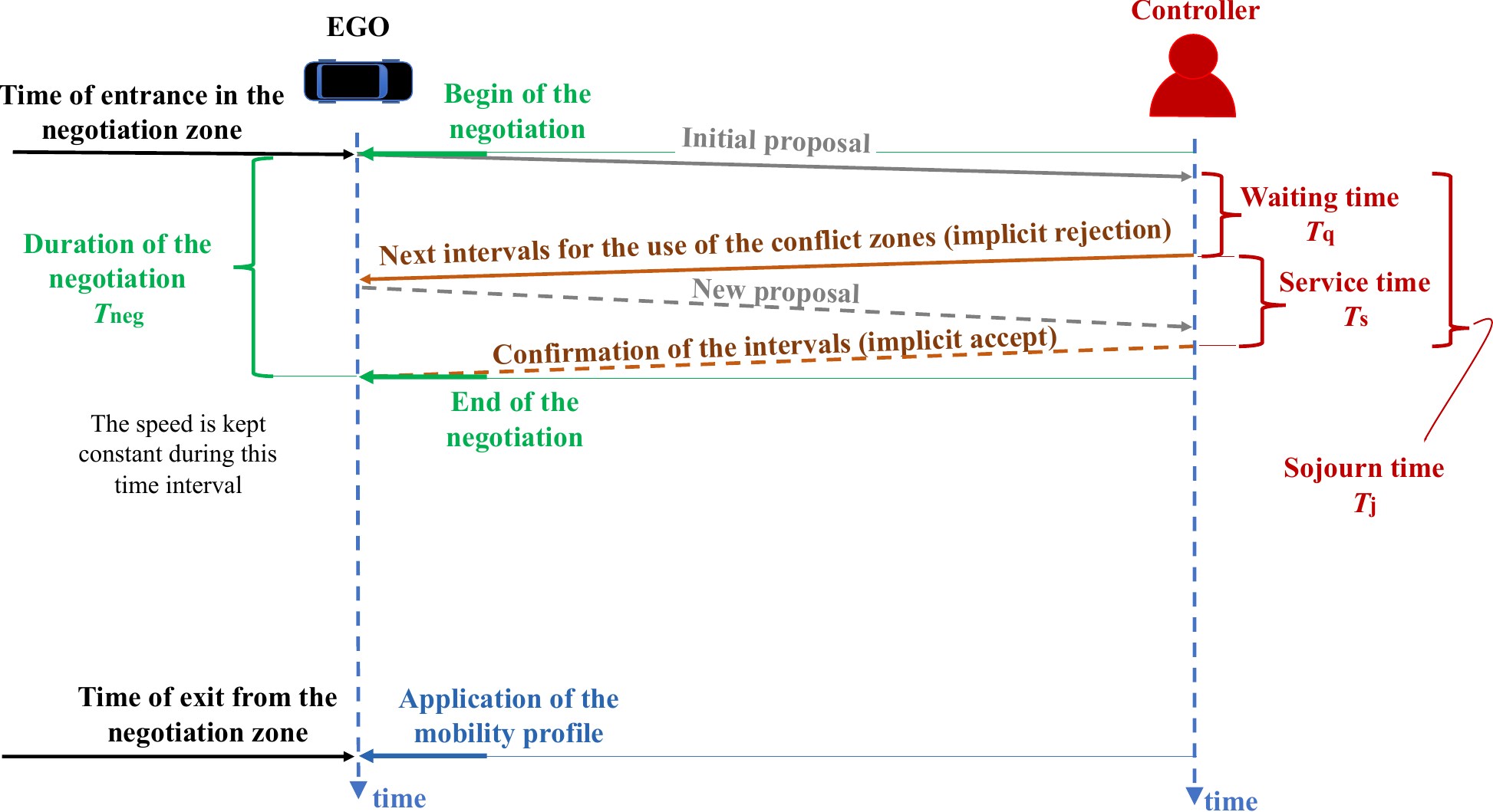}
\caption{Temporal description of the negotiation procedure, assuming as an example that the initial proposal is rejected, but the second proposal is accepted.}
\label{fig:exchange}
\end{figure*}

\subsection{Backup mode}
\label{sec:backup_mode_section}

If either: 1) the EGO exits the negotiation area before concluding the procedure; or 2) the calculated mobility profile requires a speed lower than a predefined minimum threshold, then the negotiation fails. In such cases, the EGO transmits a negotiation failure message, and the intersection management system switches to the backup mode. 

The backup mode corresponds to a critical situation where CAVs cannot all cross the intersection without stopping, and where the proposed algorithm cannot be used. The problem shifts from an optimization of the mobility profiles to the coordination of the priorities, which can be solved for example using one of the solutions proposed in \cite{BAZZI201642,JAME2022167}; in this work, we assume that in backup mode the CAVs stop coordinating and simply drive autonomously and apply legacy priorities.

\section{Analysis of the impact of communication}\label{sec:commImpact}

Before providing simulation results in Section~\ref{sec:results}, hereafter we aim at discussing the impact of communication on the correct operation of \ac{Moveover}. 

\subsection{Modeling of communication impairments}\label{sec:commmodeling} 

As detailed, the communication protocol foresees messages that are always intended for a specific target (either from the EGO to the controller or from the controller to the EGO), thus the data exchanges are implemented through unicast transmissions. Given this assumption, losses can be detected and handled with retransmissions  until they are correctly received, for example thanks to a link-layer \ac{ARQ} mechanism and a transport-layer TCP-like protocol. This architecture is compliant with the C-Roads specifications~\cite{croads-ip-based-interface,croads-cits-cross-border}, which recommend the adoption of the \ac{AMQP} v1.0 protocol for \ac{V2N} communications. Indeed, \ac{AMQP} operates over TCP at the transport layer. Therefore, the impairments in the communication can be properly reproduced as a delay, that we model as a random variable. 

In this work, we model the delay introduced for each message from the EGO to the controller and from the controller to the EGO as a random variable uniformly distributed between a minimum and a maximum delay. Different values for the minimum and maximum delay are used to model networks with different performance. In particular, the following two intervals are considered: 
\begin{itemize}
    \item 0-10 ms, reproducing a 5G network~\cite{8891450};
    \item 20-50 ms, reproducing a 4G network with LTE-Advanced technology~\cite{qualcomm_cv2x_japan_2018}; 
\end{itemize}
Despite the simplistic modeling, the statistical variability and the use of different ranges allows to derive insights of broader validity. 

\subsection{Modeling of the serving queue in the controller}\label{sec:servingqueue}

The exchanges between the EGO and controller 
are hereafter discussed. An example is also provided in Fig.~\ref{fig:exchange}, where two exchanges between the EGO and the controller happen before the negotiation is concluded. 
Upon receiving a proposal from a CAV, the controller starts its evaluation immediately if it is not engaged in a negotiation with another vehicle; otherwise, the proposal is inserted in a queue and served as soon as the controller completes the previous negotiations, following an \ac{FCFS} order. The time spent in the queue before the proposal is processed is denoted as \textit{waiting time}.
When the controller receives the first proposal, it starts its processing with possible requests of reevaluation, which end when the transmission of the last response is sent. This interval of time is denoted as \textit{service time}. 

The number of exchanges between the EGO and the server is in principle unbounded. However, in Section~\ref{sec:simulation} it will be shown that in all the simulated scenarios the large majority of the exchanges is limited to two or four, and that eight messages represent a worst case that occurs in a few cases. For this reason, hereafter we assume a conservative scenario where all negotiations require four messages to complete the procedure and a pessimistic scenario where eight messages are always needed. This approach allows us to generalize with respect to the specific intersection.  

As a consequence and assuming the model for the communication discussed in Section~\ref{sec:commmodeling}, the service time can be modeled as the sum of either two or six uniformly distributed random variables for the cases with four or eight messages, respectively. 
The resulting distribution is known as Irwin-Hall distribution. 

\begin{figure*}[t]
\centering
\begin{subfigure}[b]{0.30\textwidth}
    \includegraphics[width=\textwidth]{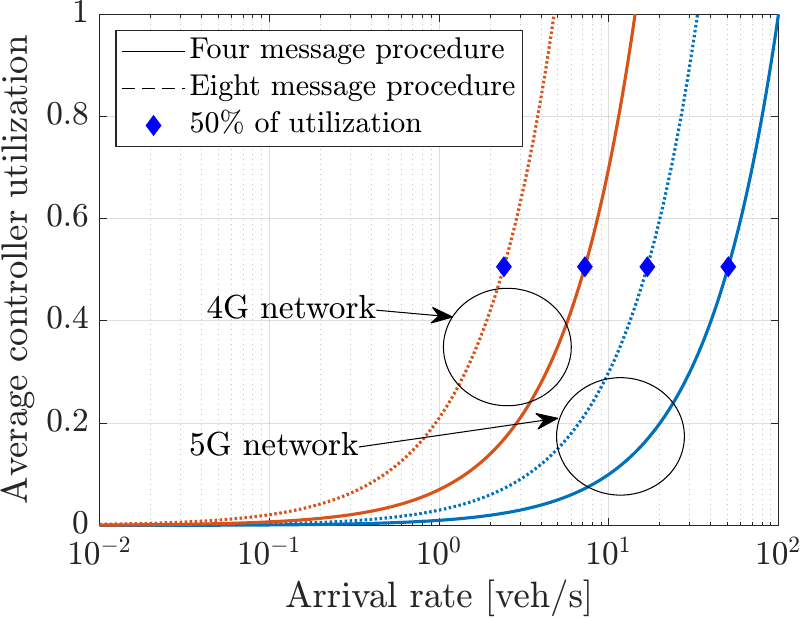}
    \caption{Average controller utilization 
    }
    \label{fig:serverUtilization}
\end{subfigure}
\hfill
\begin{subfigure}[b]{0.30\textwidth}
    \includegraphics[width=\textwidth]{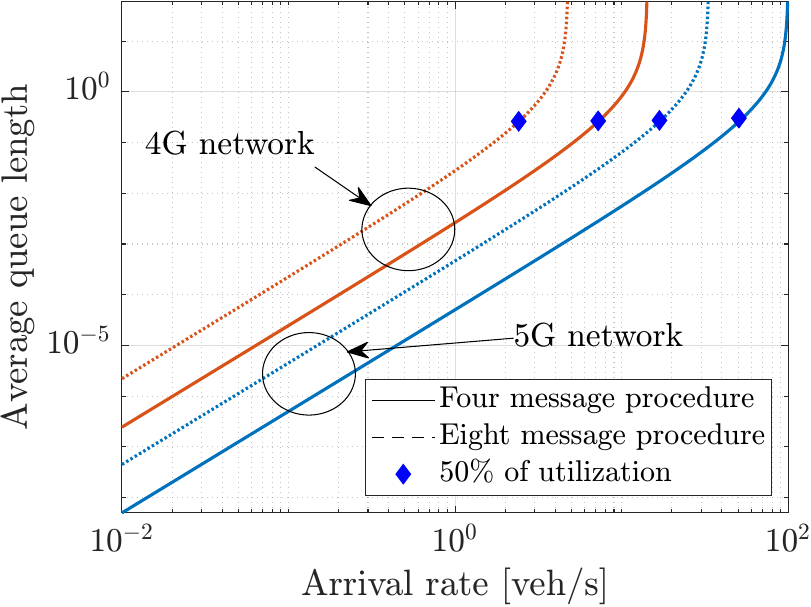}
    \caption{Average queue length}
    \label{fig:queue_length}
\end{subfigure}
\hfill
\begin{subfigure}[b]{0.30\textwidth}
    \includegraphics[width=\textwidth]{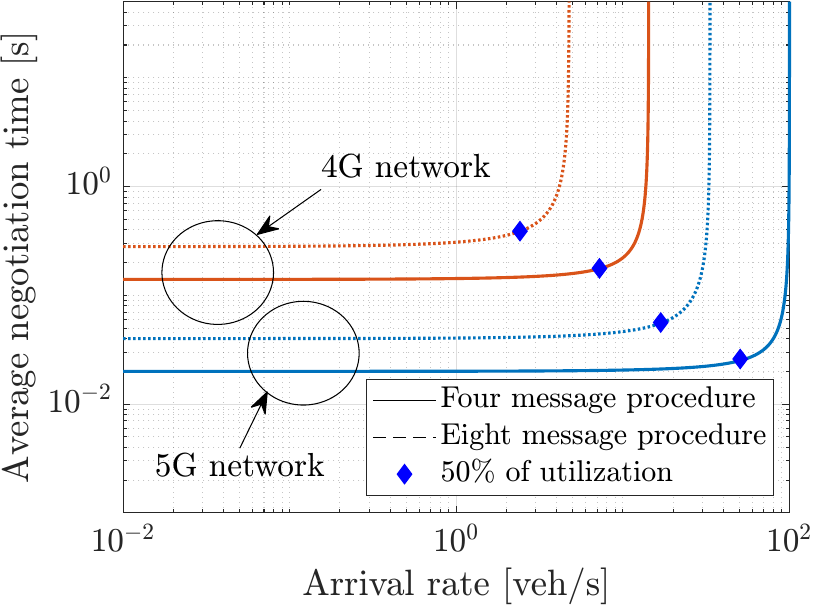}
    \caption{Average negotiation duration}
    \label{fig:sojourn_time}
\end{subfigure}
\caption{Average controller utilization, average queue length, and average negotiation duration  varying the arrival rate.}
\label{fig:queuemetrics}
\end{figure*}

\subsection{Evaluation of the impact of communication}

Following the considerations above, we model the negotiation process as an M/G/1 queue\footnote{M/G/1 queue is a model where arrivals are Markovian (modulated by a Poisson process), service times follow a distribution and there is a single server.} with an Irwin-Hall service time distribution \cite{10.5555/1972549}. In particular:

\begin{itemize}
    \item The proposal arrival is modeled as memory-less Poisson with intensity $\arrivalrate$; this is also consistent with the statistic used to generate the CAV arrival in the simulations;
    \item The service time is modeled as an Irwin-Hall distribution with mean $\servicetime = 1/\servicerate$, being $\servicerate$ the service rate, and standard deviation $\servicestd$;
    \item There is a single serving queue with infinite length.    
\end{itemize}

As a first relevant metric, the average controller utilization rate $\utilization$ can be calculated as $\utilization = \arrivalrate/\servicerate$. The utilization rate $\utilization$ is illustrated in Fig.~\ref{fig:serverUtilization} varying the arrival rate $\arrivalrate$. The plot shows that the utilization rate increases linearly with the arrival rate. 
As shown, the performance of the communication network strongly impacts system stability. Under a 4G network, where each negotiation requires hundreds of milliseconds, \ac{Moveover} can only support a few vehicles per second entering the intersection. Conversely, leveraging a 5G network reduces this delay to tens of milliseconds, allowing the system to easily manage tens of vehicles per second.

A second relevant metric is the average length of the serving queue in the controller  $\queuelength$, which can be obtained from the Little's formula \cite{10.5555/1972549} as 
\begin{equation}
\queuelength = \queuetime \cdot \arrivalrate,
\end{equation}
where $\queuetime$ is the average waiting time. 
The average queue length $\queuelength$ is shown in Fig.~\ref{fig:queue_length}, varying $\arrivalrate$ and with blue markers corresponding to a server utilization of 50\%, representing the threshold beyond which system performance begins to degrade significantly. It can be noted that when the utilization is below 50\%, the average number of requests waiting in the queue is well below 1. Then, when the utilization is close to 50\%, the queue starts exhibiting an unstable behavior. 
In such situation, the sojourn time increases significantly, some of the negotiations cannot be completed within the negotiation length, and the backup mode is entered. It is worth noting that a larger utilization of the controller corresponds to a higher density of vehicles in the intersection; in all the scenarios we have simulated, the intersection itself reaches a congested state, and the backup mode is therefore triggered, with vehicle densities that correspond to a controller utilization well below 50\%.

Finally, a relevant metric is the average duration of the negotiation $\negotiationduration$, which can be approximated as
\begin{equation}
\negotiationduration = 2 \cdot \txduration + \sojourntime = 2 \cdot \txduration + \queuetime + \servicetime\;,
\end{equation}
where: $\txduration$ is the average delay of a transmission;  
the multiplication by two times $\txduration$ accounts for the initial proposal and the final response; and $\sojourntime$ is the average \textit{sojourn time} in the system, which is composed by the average waiting time  $\queuetime$ and the average service time $\servicetime$. In turn, the average waiting time $\queuetime$ can be computed using the Pollaczek-Khinchin formula for the M/G/1 queue, as in \cite{10.5555/1972549}
\begin{equation}
  \queuetime = \arrivalrate \cdot \frac{(\frac{1}{\servicerate})^2+\servicestd^2}{2\cdot(1-\utilization)}\;,
\end{equation}
with $\servicestd^2$ being the variance of the service time obtained from the Irwin-Hall distribution. 
The average negotiation duration $\negotiationduration$ 
is depicted in Fig.~\ref{fig:sojourn_time}, again varying $\arrivalrate$ and with blue markers corresponding to an utilization of 0.5. As visible, the average negotiation duration remains almost constant until the service utilization is  approximately 0.5. We note that the negotiation duration directly impacts on the negotiation length that is required, thus relying on a faster communication allows to reduce the distance from the intersection at which the negotiation needs to be initiated. This aspect is further discussed in Appendix~\ref{sec:zonedesign}, where the design of the negotiation zone is elaborated.

\section{Numerical results}\label{sec:simulation}

The performance of the algorithm is investigated through simulations performed with the traffic simulator \ac{SUMO} \cite{SUMO2018}. When the algorithm is used, each intersection scenario is controlled by a Python script that communicates with the simulator through the \ac{TRaCI}. In particular, the script manipulates the behavior of the vehicles from the instant they enter the negotiation zone to the instant they leave the last traversed conflict zone, unless vehicles enter the backup mode. In the remaining time and during the backup mode, vehicles are directly controlled by \ac{SUMO}. Table~\ref{Tab:Settings} summarizes the main simulation settings.

\subsection{Intersection models}
We consider four different intersections, as detailed hereafter. A multi-intersection scenario has also been evaluated, as discussed in Section~\ref{sec:multiintersection}.
In all four intersections, 
the converging roads are assumed of length $2\controlLength$, where $\controlLength$=100~m is the distance from the start of the negotiation zone to the beginning of the intersection itself. The conflict zones are defined in accordance to the geometry of the considered intersection layout. The negotiation length is dependent on the intersection and is computed as the product of the maximum vehicle speed allowed inside the negotiation zone and the maximum time for the negotiation; the latter is calculated via Matlab simulations as the 99th percentile of the negotiation durations, assuming a worst-case scenario where all vehicles always exchange the maximum number of messages at the maximum delay per message and it is obtained from \ac{SUMO} simulations. We detail the four type of intersections hereafter.

\begin{figure*}[t]
\centering
\subfloat[]{\includegraphics[width=0.48\columnwidth]{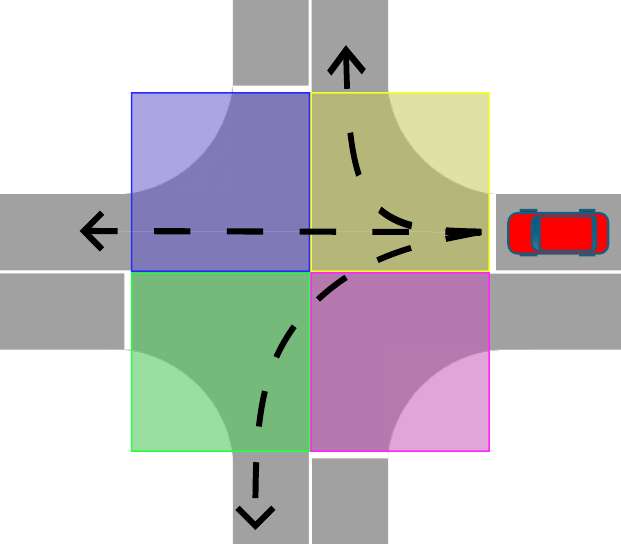}
\label{fig:exampledirections_a}}
\hfill
\subfloat[]{\includegraphics[width=0.48\columnwidth]{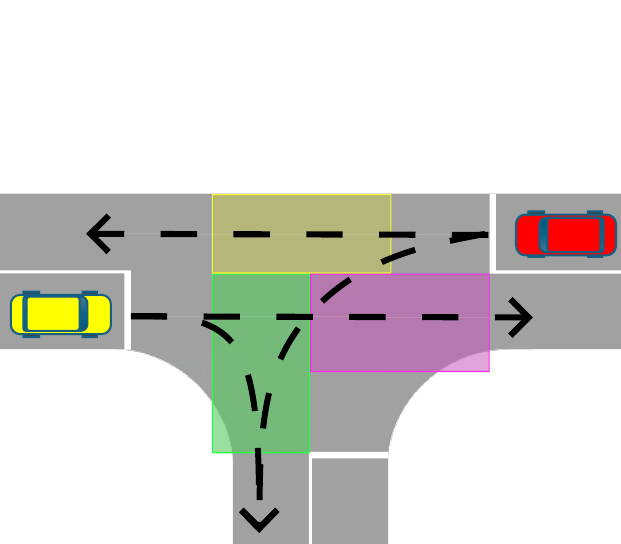}
\label{fig:exampledirections_b}}
\hfill
\subfloat[]{\includegraphics[width=0.48\columnwidth]{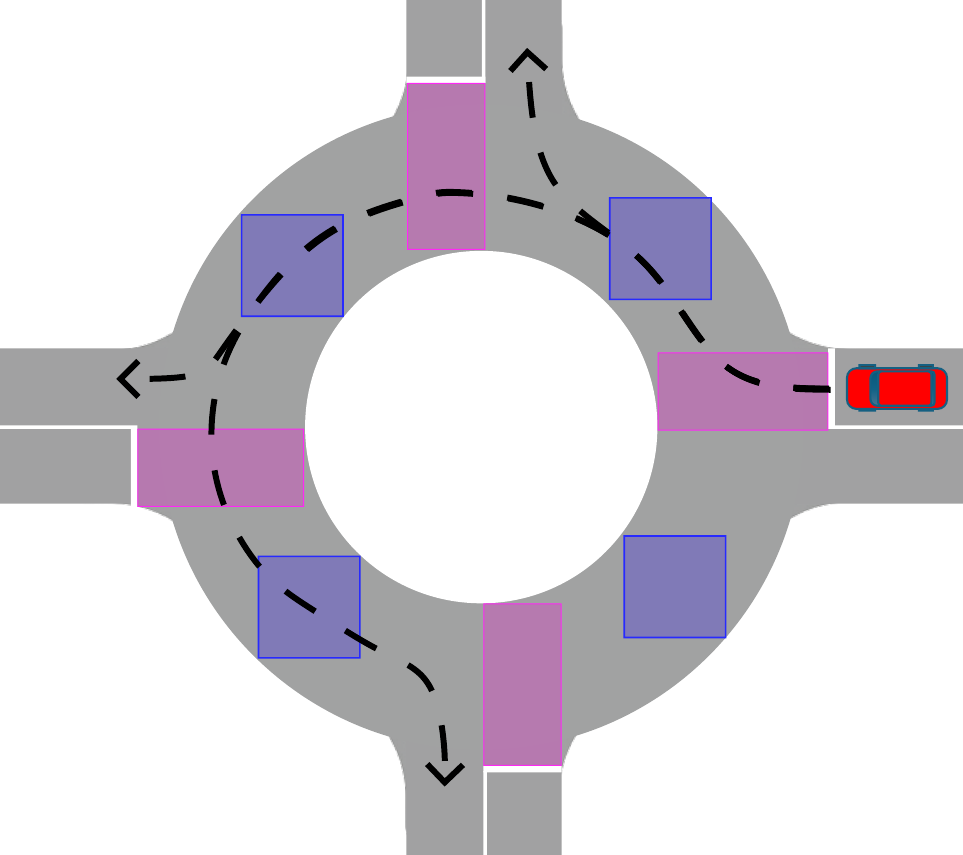}
\label{fig:exampledirections_c}}
\hfill
\subfloat[]{\includegraphics[width=0.48\columnwidth]{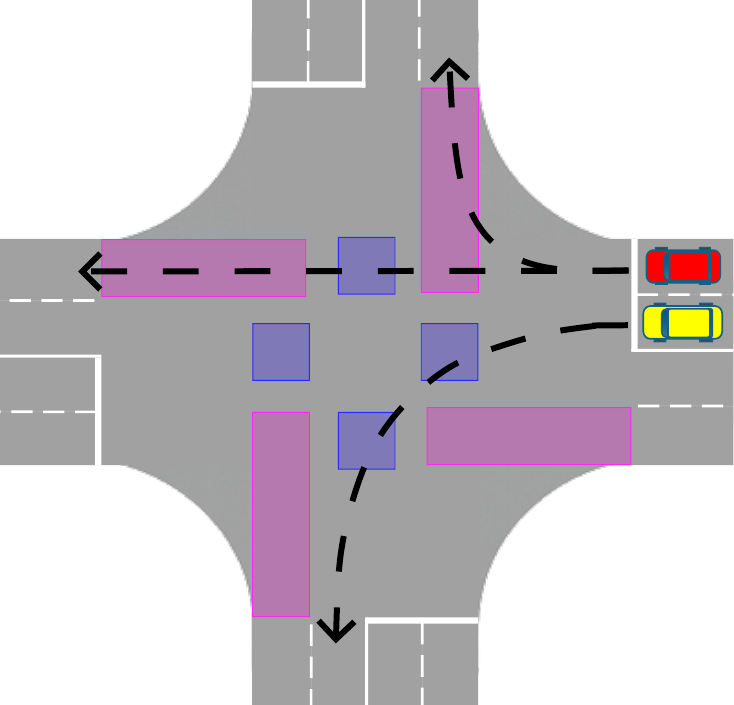}
\label{fig:exampledirections_d}}
\caption{Intersections with conflict zones and possible trajectories of the EGO. (a) Four-way single-lane intersection. (b) Three-way single-lane intersection. (c) Roundabout. (d) Four-way two-lane intersection.}
\label{fig:model2}
\end{figure*}


\begin{table}[t]
\caption{Main simulation parameters and settings.
}\label{Tab:Settings}
\footnotesize
\centering
\begin{tabular}{p{5.0cm}p{3.0cm}}
\hline \hline
\multicolumn{2}{c}{\textbf{Common simulation settings}} \\
\hline \hline
\textbf{Parameter} & \textbf{Value} \\  
Duration of each simulation & 7200 s\\
Lane width & 3.2 m \\
Distance between the start of the negotiation zone and the intersection $\controlLength$ & 100 m\\
Road length $2 \controlLength$ & 200 m\\
Vehicle arrival distribution & Poisson, variable avg.\\
Vehicle length & 5 m \\
Vehicle width & 1.8 m \\
Maximum acceleration $\accelerationmax$ & 2.6 m/s$^2$\\
Maximum braking $\accelerationmin$ & -4.5 m/s$^2$\\
Safe margin after the intersection & 6 m\\
\hline \hline
\multicolumn{2}{c}{\textbf{Four-way single-lane intersection settings}} \\
\hline \hline
Road layout & 4 roads, one lane each \\
Square conflict zone side & 7.2 m\\
Maximum speed $\speedmax$ & 50 km/h \\ 
Maximum turn speed $\speedturn$ & 20 km/h \\
Negotiation zone length (5G, 4G) & 2 m, 10 m \\
\hline \hline
\multicolumn{2}{c}{\textbf{Three-way single-lane intersection settings}} \\
\hline \hline
Road layout & 3 roads, one lane each \\
Rectangular conflict zone (length, width) & 7.2 m, 3.6 m\\
Maximum speed $\speedmax$ & 50 km/h \\ 
Maximum turn speed $\speedturn$ & 20 km/h \\
Negotiation zone length (5G, 4G) & 2 m, 10 m \\
\hline \hline
\multicolumn{2}{c}{\textbf{Roundabout settings}} \\
\hline \hline
Road layout & 4 roads, one lane each \\
Rectangular conflict zone (length, width) & 5.2 m, 3.2 m\\
Square conflict zone side & 3 m \\
Maximum speed $\speedmax$ & 23 km/h \\
Negotiation zone length (5G, 4G) & 2 m, 7.5 m \\
\hline \hline
\multicolumn{2}{c}{\textbf{Four-way two-lane intersection settings}} \\
\hline \hline
Road layout & 4 roads, two lane each \\
Rectangular conflict zone (length, width) & 8.2 m, 3.2 m \\
Square conflict zone side & 3.2 m\\
Maximum speed $\speedmax$ & 50 km/h \\ 
Maximum turn speed $\speedturn$ & 20 km/h \\
Negotiation zone length (5G, 4G) & 2 m, 17 m \\
\hline \hline
\end{tabular}
\end{table}

\subsubsection{Four-way single-lane intersection}

The junction is fully covered by four equal square conflict zones with side $7.2$~m. As represented in Fig.~\ref{fig:exampledirections_a}, the vehicle crosses one zone when turning right, two zones when going straight, and three zones when turning left. The maximum number of exchanged messages is 8, resulting in a negotiation length of $2$~m for the 5G network and $10$~m for the 4G network. 

\subsubsection{Three-way single-lane intersection}
Three rectangular conflict zones are designed with length $7.2$~m and width $3.6$~m. As shown in Fig.~\ref{fig:exampledirections_b}, a vehicle always crosses one zone when turning right and three zones when turning left. The number of zones crossed when going straight depends on the arrival road. In this scenario, the number of exchanged messages is again at most 8, leading to a negotiation length of $2$~m and $10$~m for the 5G and 4G networks, respectively. 

\subsubsection{Roundabout}
Eight conflict zones of two types are identified on the roundabout: four rectangular zones of length $5.2$~m and width $3.2$~m at the four entrances of the roundabout, and four square zones of side $3$~m halfway between every entrance and the respective first exit. The possible trajectories are illustrated in Fig.~\ref{fig:exampledirections_c}: the vehicle crosses two zones when leaving at the first exit, four zones when leaving at the second exit, and six zones when leaving at the third exit. At most 6 messages are exchanged between vehicles and the controller; thus, the computed negotiation length is $2$~m for the 5G network and $7.5$~m for the 4G network.

\subsubsection{Four-way two-lane intersection}
Also in this case, eight conflict zones of two types are designed: four rectangular zones with length $8.2$~m and width $3.2$~m, and four square zones with side $3.2$~m. The vehicle crosses one zone when turning right, two zones when turning left, and three zones when going straight. These trajectories are represented in Fig.~\ref{fig:exampledirections_d}. When assuming the presence of a 5G network, at most 8 messages are exchanged, leading to a negotiation length of $2$~m; with a 4G network, the maximum number of messages increases to 10, requiring a negotiation length of $17$~m.

\begin{figure*}[h]
\centering
\subfloat[]
{\includegraphics[width=0.9\columnwidth]{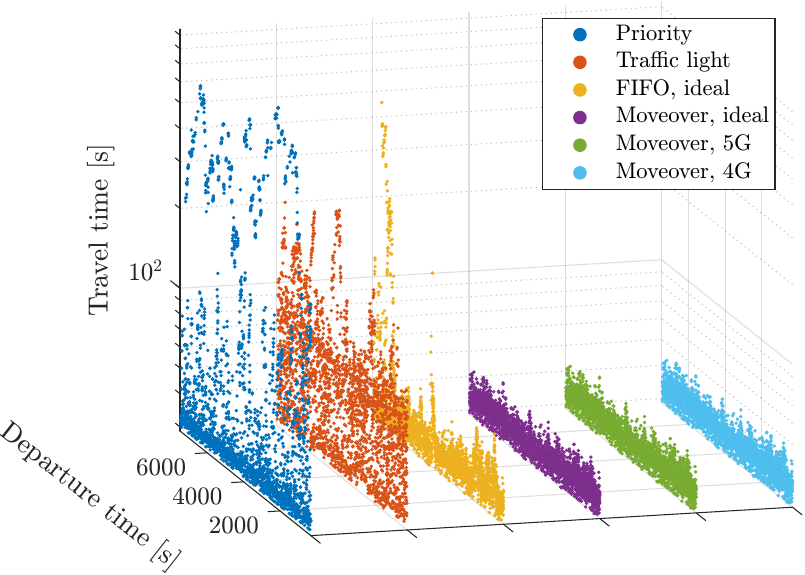}
\label{fig:Traveltime_fourway}}
\subfloat[]
{\includegraphics[width=0.9\columnwidth]{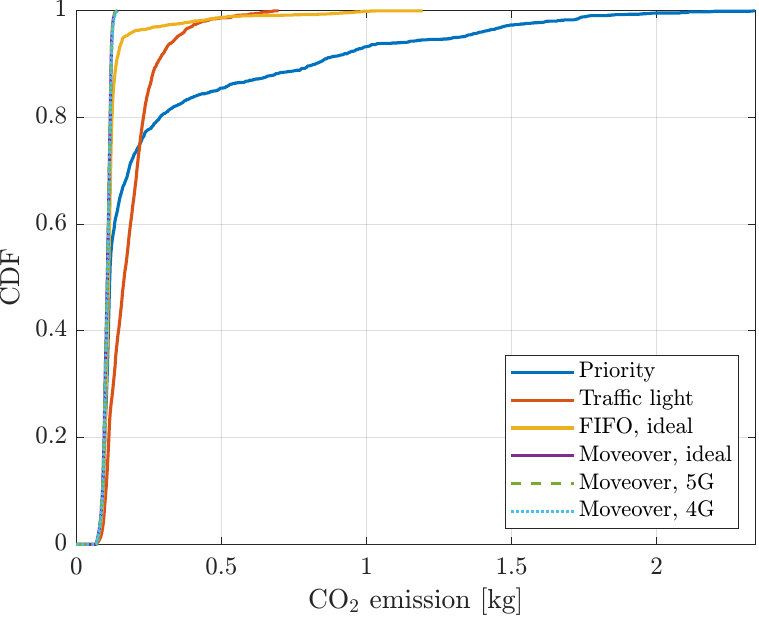}
\label{fig:Emission_fourway}}
\caption{Results for the four-way single-lane scenario with 0.1 vehicles/s on average per each direction. (a) Travel time over departure time. (b) Cdf of the CO2 emissions.}
\label{fig:fourway}
\end{figure*}

\subsection{Benchmarks}
We consider and compare the following approaches, which possibly involve either \textit{ideal}, 5G, or 4G communication. In particular, we refer to ideal communication to consider the case where messages are always correctly delivered with negligible delay, meaning that the decisions are instantaneously taken by the controller. In such case, the negotiation length is set to $0$~m. Otherwise, for 5G and 4G, we take into account delays discussed in Section~\ref{sec:commmodeling}.
The following approaches are compared:
\begin{itemize}
    \item \textit{Priority} approach, where vehicles on horizontal lanes have precedence over vertical ones;
    \item \textit{Traffic light};
    \item \textit{FIFO} \cite{zhang2019decentralized} algorithm with ideal communication;
    \item \textit{\ac{Moveover}} with ideal communication; 
    \item \textit{\ac{Moveover}} with 5G connectivity;
    \item \textit{\ac{Moveover}} with 4G connectivity;
\end{itemize}

\begin{rmk}
Each traffic light has a green phase of 35 seconds and a yellow phase of 3 seconds for streets facing each other in the four-way single-lane scenario; the same timing is applied in the three-way single-lane scenario. In the four-way two-lane intersection, the traffic light provides a green phase of 33 seconds followed by a yellow phase of 3 seconds for the straight and right directions. The left direction remains green for an additional 9 seconds, followed by a yellow phase of 3 seconds.
\end{rmk}

\begin{rmk}
In the roundabout scenario, the priority benchmark refers to the roundabout operating under normal conditions. In the same scenario, the traffic light benchmark is not considered, as roundabouts are not commonly managed with traffic lights, while the FIFO benchmark is excluded as it leads to unmanageable traffic congestion even at low vehicle densities.
\end{rmk}

\begin{figure*}[h]
\centering
\subfloat[]
{\includegraphics[width=0.9\columnwidth]{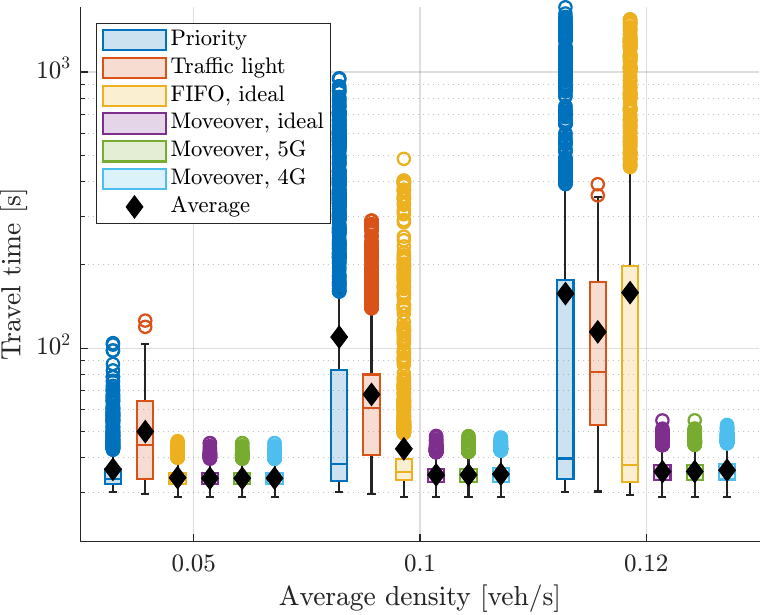}
\label{fig:Traveltimebc_fourway}}
\subfloat[]{\includegraphics[width=0.9\columnwidth]{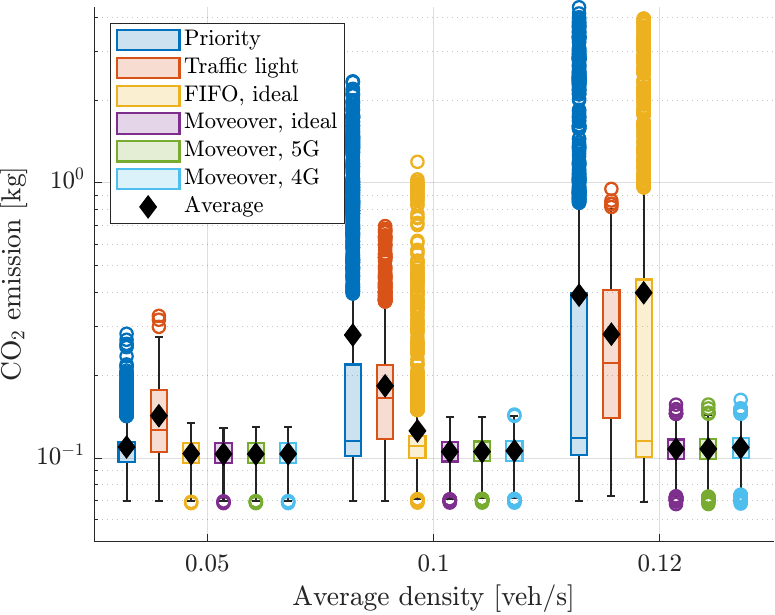}
\label{fig:Emissionbc_fourway}}
\caption{Box charts for the four-way single-lane scenario varying the density. (a) Travel time. (b) CO2 Emissions.}
\label{fig:boxchart_fourway}
\end{figure*}

\begin{figure*}[t]
\centering
\subfloat[]
{\includegraphics[width=0.9\columnwidth]{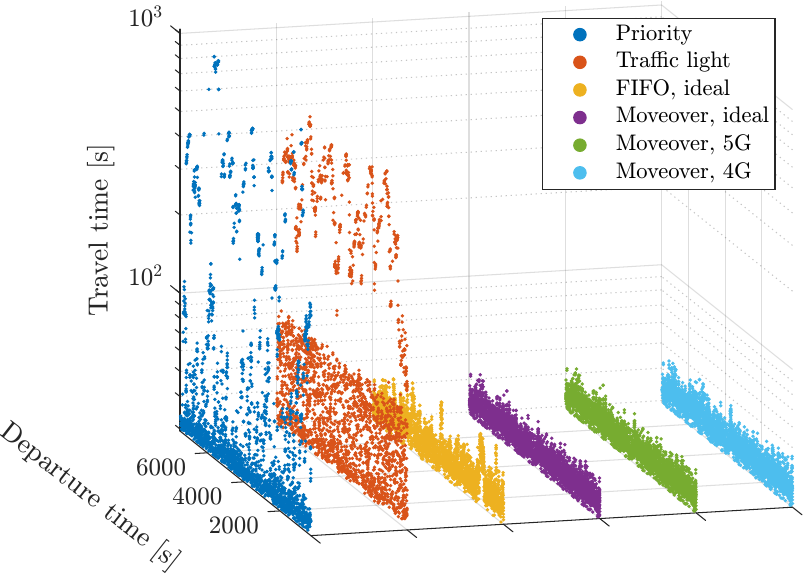}
\label{fig:Traveltime_threeway}}
\subfloat[]
{\includegraphics[width=0.9\columnwidth]{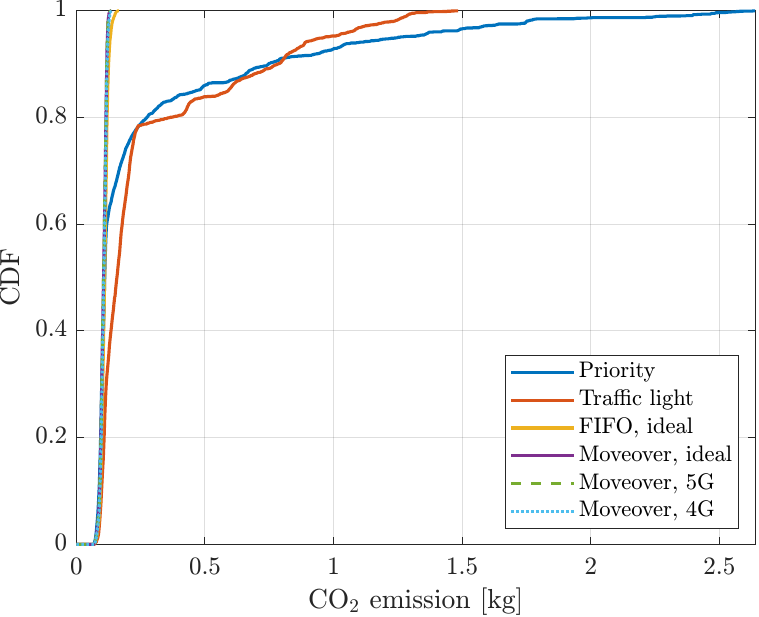}
\label{fig:Emission_threeway}}
\caption{Results for the three-way single-lane scenario with 0.15 vehicles/s on average per each direction. (a) Travel time over departure time. (b) Cdf of the CO2 emissions.}
\label{fig:threeway}
\end{figure*}

\subsection{Evaluation metrics}
Performance is assessed through the following \acp{KPI}:

\begin{itemize}
    \item \textbf{Travel time}: time elapsed in seconds from the instant a vehicle enters the simulation to the instant it leaves;
    \item \textbf{Emissions}: CO$_2$ emissions from vehicles during their travel time, assuming the \texttt{HBEFA3/PC\_G\_EU4} SUMO emission class. This model characterizes the emissions of a gasoline-powered Euro 4 passenger car according to the Handbook Emission Factors for Road Transport (HBEFA) version 3~\cite{hbefa2010}.
\end{itemize}

\subsection{Results with single intersections}\label{sec:results}

\begin{figure*}[t]
\centering
\subfloat[]
{\includegraphics[width=0.9\columnwidth]{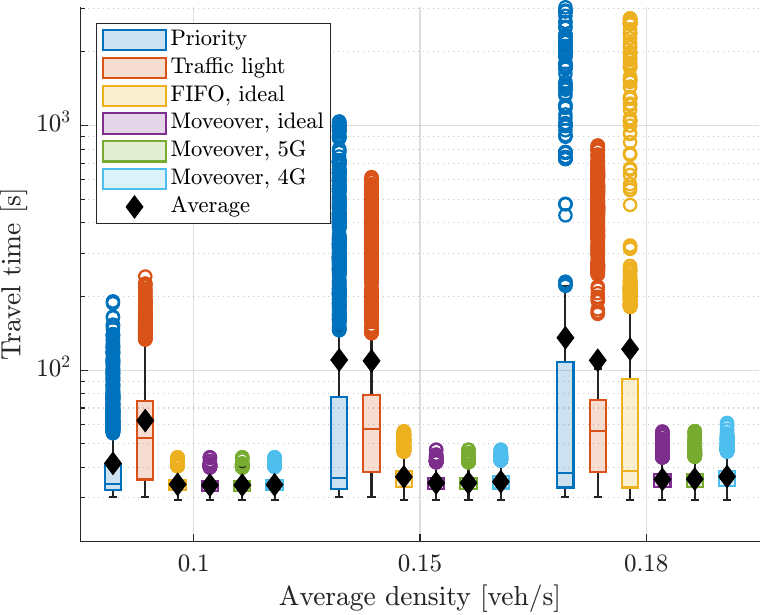}
\label{fig:Traveltimebc_threeway}}
\subfloat[]{\includegraphics[width=0.9\columnwidth]{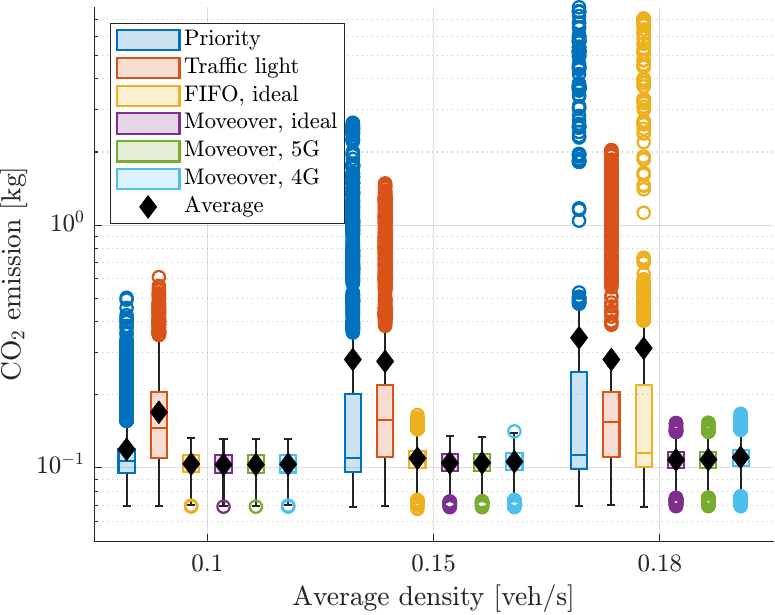}
\label{fig:Emissionbc_threeway}}
\caption{Box charts for the three-way single-lane scenario varying the density. (a) Travel time. (b) CO2 emissions.}a   
\label{fig:boxchart_threeway}
\end{figure*}

\begin{figure*}[t]
\centering
\subfloat[]
{\includegraphics[width=0.9\columnwidth]{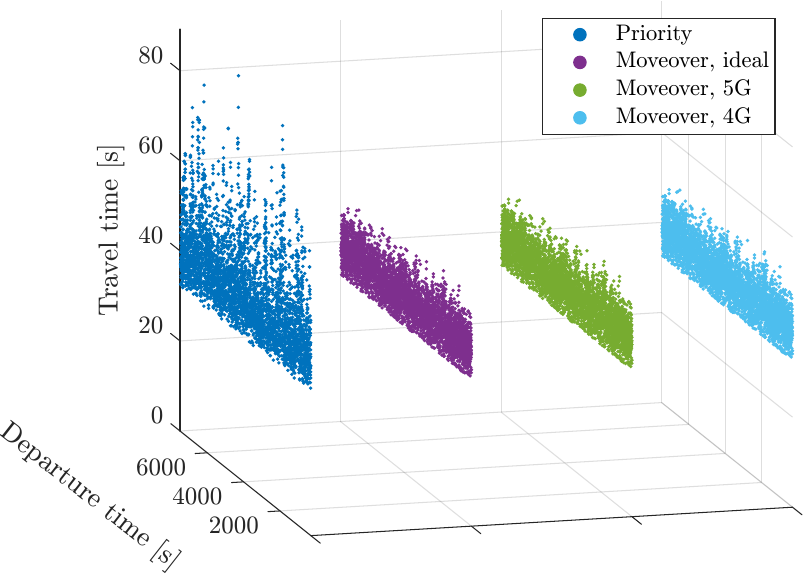}
\label{fig:Traveltime_roundabout}}
\subfloat[]
{\includegraphics[width=0.9\columnwidth]{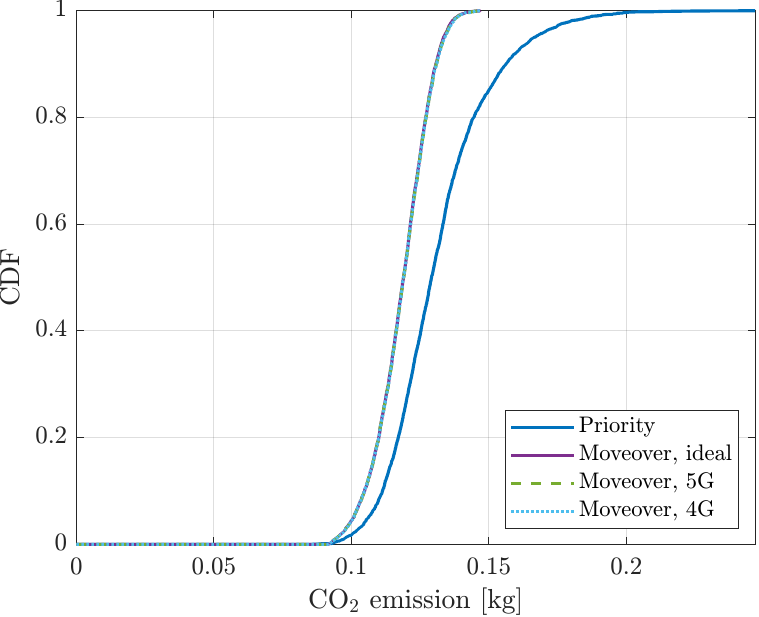}
\label{fig:Emission_roundabout}}
\caption{Results for the roundabout scenario with 0.15 vehicles/s on average per each direction. (a) Travel time over departure time. (b) Cdf of the CO2 emissions.}
\label{fig:roundabout}
\end{figure*}

For each scenario, four different plots are depicted: a graph showing travel time versus departure time and the \ac{cdf} of the CO$_2$ emissions at a given vehicle density, together with box charts of travel times and emissions for varying traffic densities. In addition to Figs.~\ref{fig:fourway}-\ref{fig:boxchart_fourwaytwo}, which are related to each single scenario, Figs.~\ref{fig:density}, ~\ref{fig:messages_5G}, and~\ref{fig:messages_4G} provide summary results on the intersection capacity and the number of exchanged messages.

\subsubsection{Four-way single-lane intersection}

Results for the four-way single lane intersection are shown in Figs.~\ref{fig:fourway} and \ref{fig:boxchart_fourway}. Fig.~\ref{fig:Traveltime_fourway} represents the travel time versus the departure time for an average vehicle density of 0.1 \ac{veh/s} per direction. Here, the priority case exhibits the worst performance, with several vehicles, specifically those traveling on the road without priority, experiencing large travel times. The traffic light-controlled scenario also results in high travel times, which progressively increases during the simulation. In contrast, the FIFO-based approach achieves good performance for most of the simulation; however, a sharp increase in travel times towards the end indicates the onset of congestion. The proposed algorithm is able to outperform all the considered benchmarks, maintaining improved results throughout the whole simulation. Moreover, it proves resilient to communication delays, achieving a satisfactory performance under both 5G and 4G network conditions.
Fig.~\ref{fig:Emission_fourway} instead illustrates the \ac{cdf} of the CO$_2$ emissions at the same traffic density. The observed trends are consistent with those discussed for travel time, reflecting the strong correlation between congestion levels and emissions production. Accordingly, the proposed algorithm again exhibits clear advantages over the benchmarks.

Fig.~\ref{fig:boxchart_fourway} depicts box charts of the same two metrics for varying vehicle densities per direction. It can be noted that, at low densities, the priority-based approach performs better than the traffic light-controlled scenario, as the latter introduces unnecessary stops and increased emissions. However, as density increases, the traffic light strategy begins to outperform the priority approach. The FIFO-based solution maintains acceptable performance up to medium densities, but deteriorates under high traffic volumes. In contrast, \ac{Moveover} achieves steady and robust results across all considered densities.

\subsubsection{Three-way single-lane intersection}

Figs.~\ref{fig:threeway} and \ref{fig:boxchart_threeway} depict the results for the three-way single-lane scenario. Fig.~\ref{fig:Traveltime_threeway} illustrates the travel time versus the departure time, while Fig.~\ref{fig:Emission_threeway} shows the \ac{cdf} of the CO$_2$ emissions, both at an average vehicle density of 0.15~\ac{veh/s} per direction. Regarding the priority-based and the traffic light-based approaches, the results are similar to those of the four-way single-lane intersection. On the contrary, the FIFO-based solution maintains an acceptable performance for the entirety of the simulation, with results comparable to those of the proposed algorithm. 

However, when the density increases, all three benchmarks undergo an evident degradation of the performance, as shown with box charts in Figs.~\ref{fig:Traveltimebc_threeway} and \ref{fig:Emissionbc_threeway}. Similarly to the previously discussed scenario, the traffic light-strategy starts yielding benefits compared to the priority case when increasing the traffic density. The FIFO approach proves to be more resilient, but it still strongly degrades at the highest density. \ac{Moveover} instead guarantees an adequate performance at all considered traffic volumes, and the degradation caused by communication delays is still negligible.   

\subsubsection{Roundabout}

\begin{figure*}[t]
\centering
\subfloat[]
{\includegraphics[width=0.9\columnwidth]{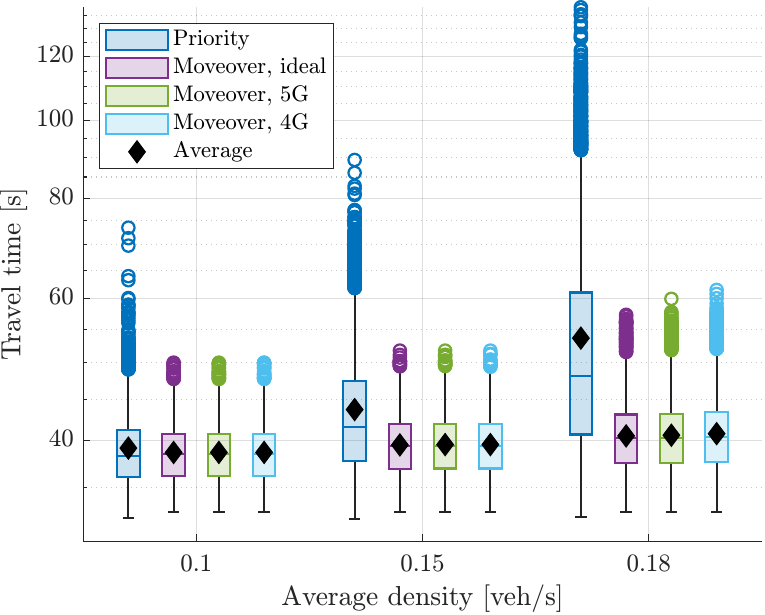}
\label{fig:Traveltimebc_roundabout}}
\subfloat[]{\includegraphics[width=0.9\columnwidth]{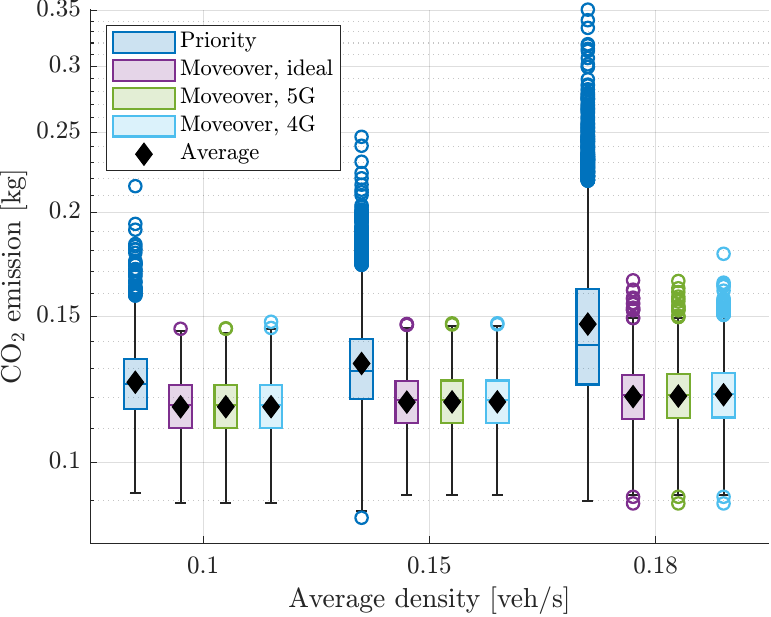}
\label{fig:Emissionbc_roundabout}}
\caption{Box charts for the roundabout scenario varying the density. (a) Travel time. (b) CO2 emissions.}
\label{fig:boxchart_roundabout}
\end{figure*}

Results for the roundabout are shown in Figs.~\ref{fig:roundabout} and \ref{fig:boxchart_roundabout}. 
Travel time over departure time and the \ac{cdf} of the CO$_2$ emissions at an average traffic density of 0.15~\ac{veh/s} per direction are shown in Figs.~\ref{fig:Traveltime_roundabout} and \ref{fig:Emission_roundabout}, respectively. It can be noted that the performance gap between \ac{Moveover} and the priority-based approach is less significant than in the scenarios analyzed earlier. This stems from the fact that roundabouts possess a higher inherent traffic efficiency compared to conventional intersections. Nevertheless, the impact of the proposed solution scales with the traffic load, as demonstrated by the box charts in Fig.~\ref{fig:boxchart_roundabout}. Indeed, the benefits in terms of both metrics become more evident as the vehicle density increases. Furthermore, the sensitivity of the algorithm to communication delay remains negligible, even under demanding traffic conditions.

\begin{figure*}[t]
\centering
\subfloat[]
{\includegraphics[width=0.9\columnwidth]{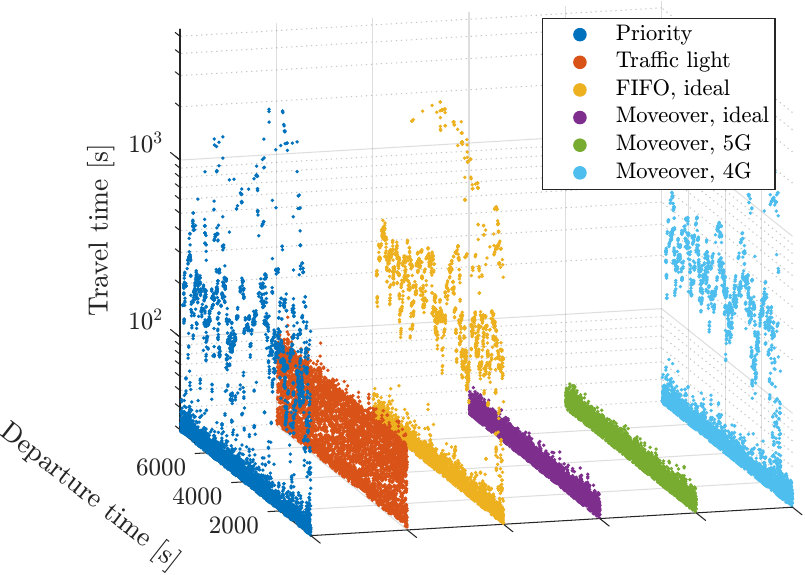}
\label{fig:Traveltime_fourwaytwo}}
\subfloat[]
{\includegraphics[width=0.9\columnwidth]{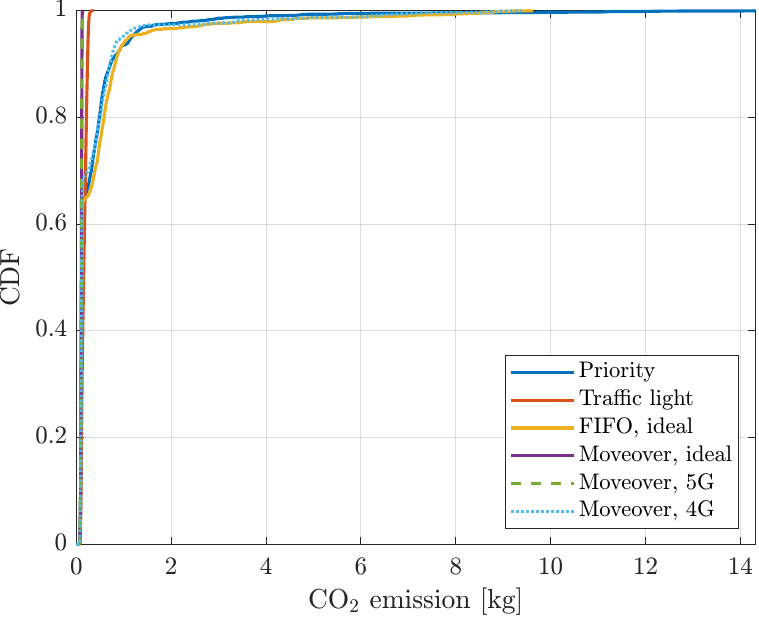}
\label{fig:Emission_fourwaytwo}}
\caption{Results for the four-way two-lane scenario with 0.2 vehicles/s on average per each direction. (a) Travel time over departure time. (b) Cdf of the CO2 emissions.}
\label{fig:fourwaytwo}
\end{figure*}

\subsubsection{Four-way two-lane intersection}

\begin{figure*}[t]
\centering
\subfloat[]
{\includegraphics[width=0.9\columnwidth]{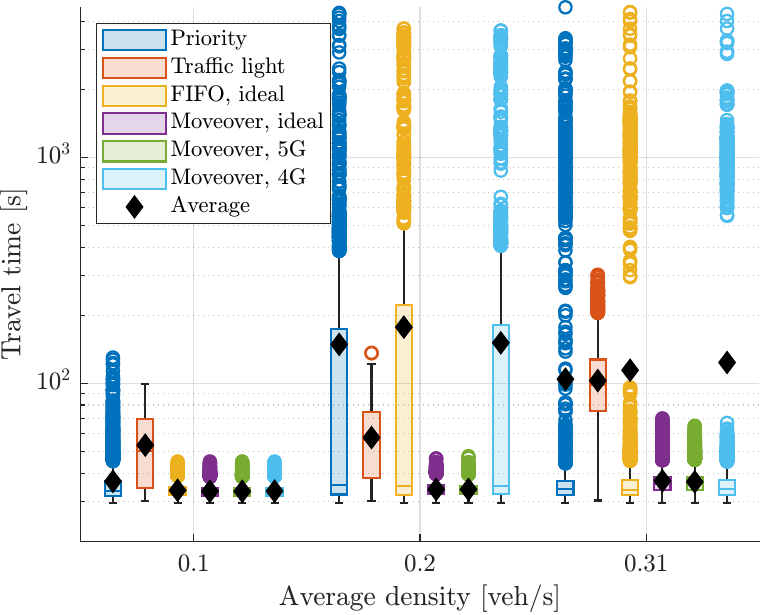}
\label{fig:Traveltimebc_fourwaytwo}}
\subfloat[]{\includegraphics[width=0.9\columnwidth]{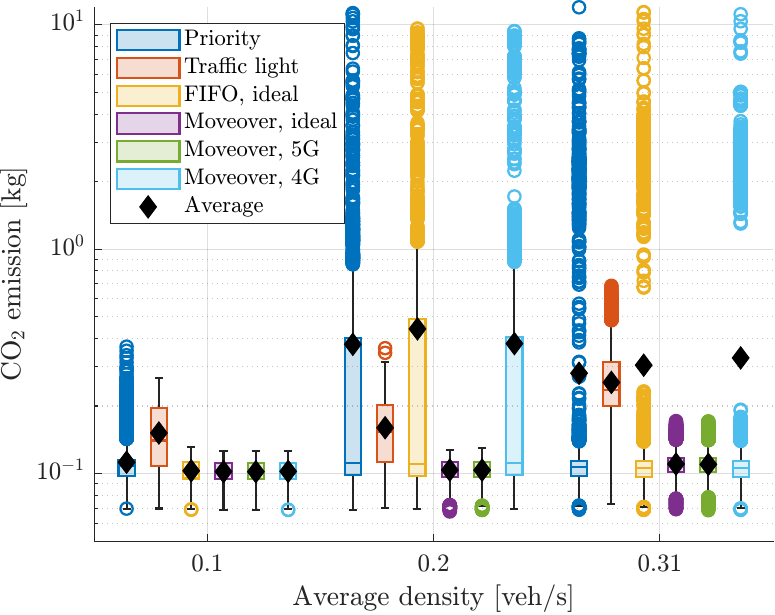}
\label{fig:Emissionbc_fourwaytwo}}
\caption{Box charts for the four-way two-lane scenario varying the density. (a) Travel time. (b) CO2 emissions.}
\label{fig:boxchart_fourwaytwo}
\end{figure*}

Figs.~\ref{fig:fourwaytwo} and \ref{fig:boxchart_fourwaytwo} report the results for the four-way two-lane intersection. Specific insights into this scenario can be inferred by observing the travel time versus departure time and the \ac{cdf} of CO$_2$ emissions in Figs.~\ref{fig:Traveltime_fourwaytwo} and \ref{fig:Emission_fourwaytwo}, both recorded at an average traffic density of 0.2~\ac{veh/s} per direction. The priority-based and the FIFO-based approaches are evidently unable to manage the considered traffic density, with strong peaks of travel time and emissions. The traffic light case proves to be the best performing benchmark among the first three, still providing a decent performance. Finally, \ac{Moveover} clearly outperforms the other traffic management methods. Nevertheless, while the 5G-based architecture remains robust to communication impairments, the 4G network case exhibits a significant performance degradation. This behavior is to be attributed to the required length of the negotiation zone: to ensure reliable service for all users in this scenario, the zone must be extended to 17~meters. Such a distance, however, promotes the onset of congestion, triggering the backup mode at lower vehicle densities compared to the 5G case.

The performance evolution of the various approaches under varying traffic volumes is depicted in Fig.~\ref{fig:boxchart_fourwaytwo}. Consistently with the four-way single-lane scenario, the traffic light case is the least effective at low densities; however, as the density increases, it outperforms both the priority-based and FIFO-based approaches, which instead undergo severe degradation. \ac{Moveover} yields the greatest benefits under the 5G network assumption across the whole density range, whereas the 4G-based architecture fails to sustain medium-to-high vehicle densities.

\begin{figure}[t]
\centering
\includegraphics[width=0.8\columnwidth]{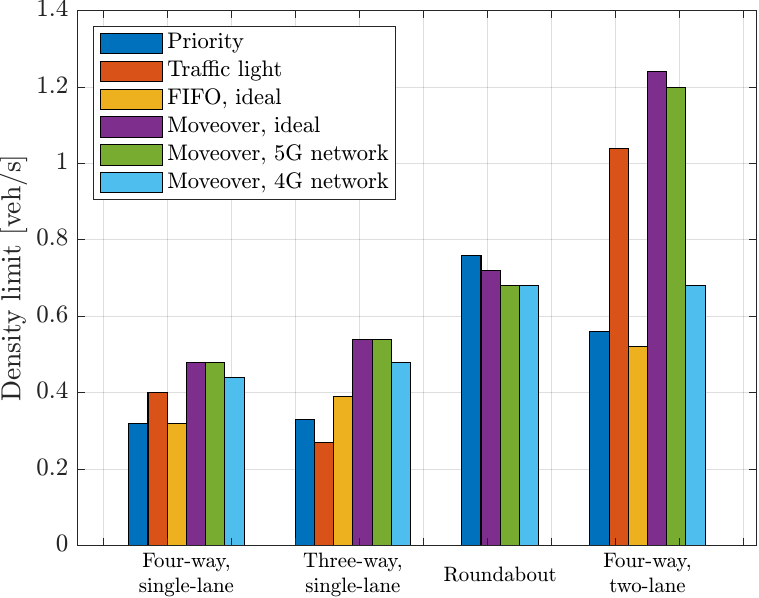}
\caption{Maximum densities for the different scenarios and traffic management methods.}
\label{fig:density}
\end{figure}

\begin{figure}[t]
\centering
\includegraphics[width=0.8\columnwidth]{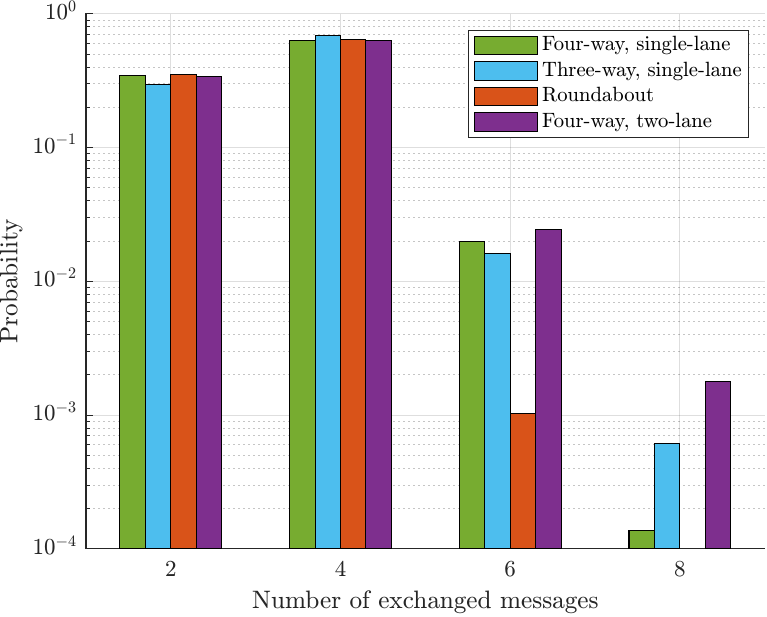}
\caption{Empirical probability distribution of exchanged messages at the maximum sustainable densities for different scenarios (5G network).}
\label{fig:messages_5G}
\end{figure}

\begin{figure}[t]
\centering
\includegraphics[width=0.8\columnwidth]{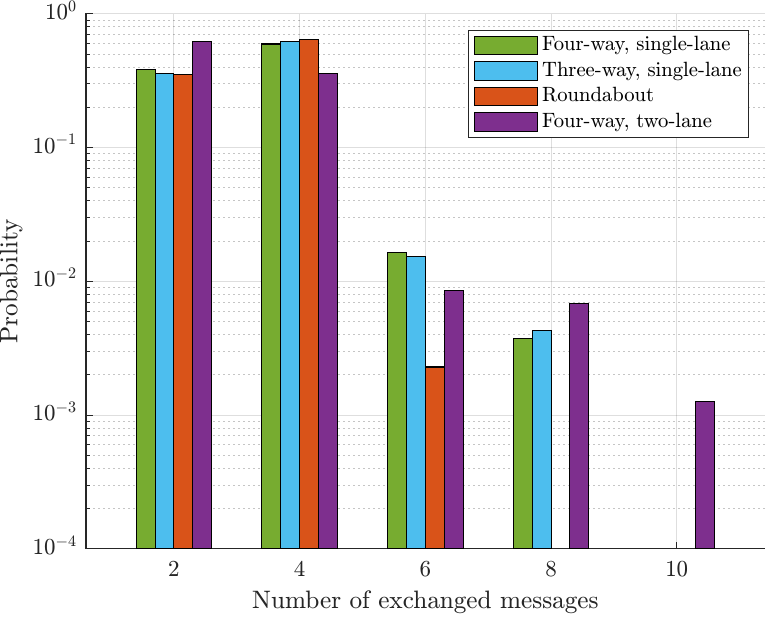}
\caption{Empirical probability distribution of exchanged messages at the maximum sustainable densities for different scenarios (4G network).}
\label{fig:messages_4G}
\end{figure}

\begin{figure}[t]
\centering
\subfloat{\includegraphics[trim=0cm 3cm 5cm 3cm, clip, width=0.7\columnwidth]{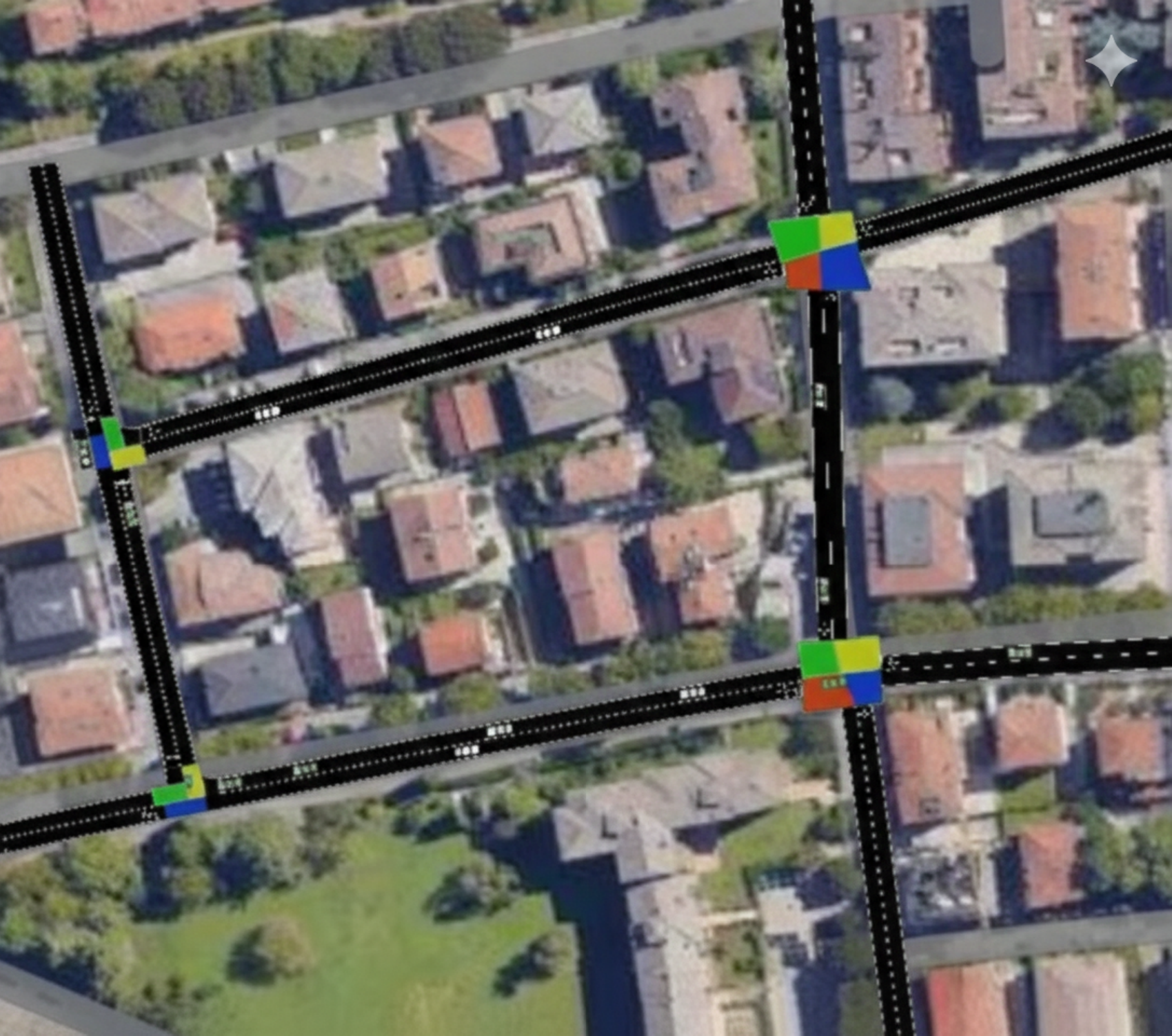}}
\caption{Suburban area with four controlled intersections and the conflict zones of each intersection.}
\label{fig:openstreetmapscenario}
\vskip -0.3cm
\end{figure}

\subsubsection{Comparative analysis of the scenarios}

\begin{figure*}[t]
\centering
\subfloat[]
{\includegraphics[width=0.9\columnwidth]{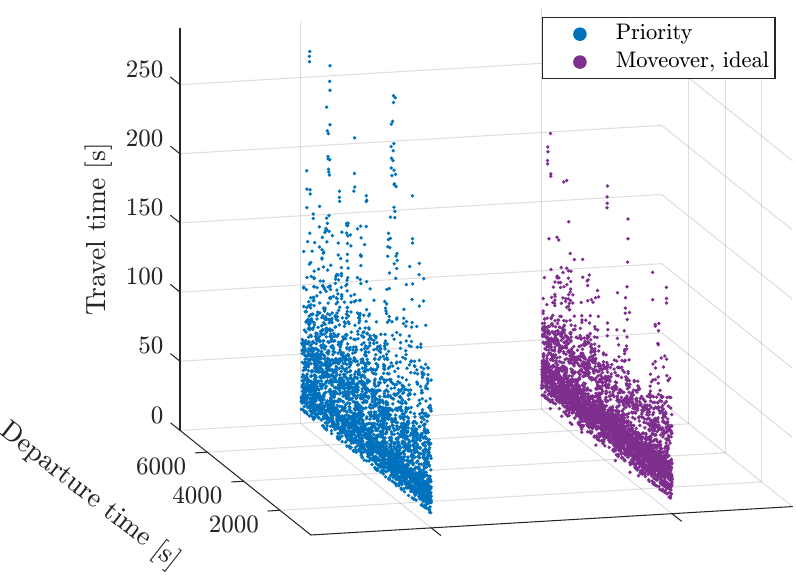}
\label{fig:Traveltime_multi}}
\subfloat[]
{\includegraphics[width=0.9\columnwidth]{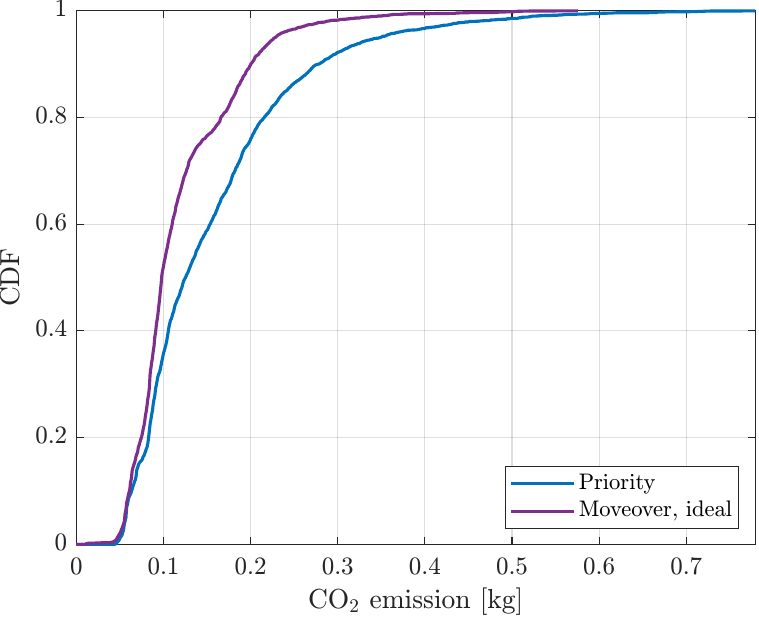}
\label{fig:Emission_multi}}
\caption{Results for the real urban map with 0.5 vehicles/s. (a) Travel time over departure time. (b) Cdf of the CO2 emissions.}
\label{fig:multiNetwork}
\end{figure*}

Fig.~\ref{fig:density} summarizes the maximum sustainable vehicle density achieved by each traffic management method across the considered scenarios. A traffic management method is considered sustainable at a specific density level if its 90th percentile of the travel times remains below a a scenario-specific threshold, which is computed as three times the average travel time recorded under the priority-based approach at a low vehicle density.
Analyzing the plot, several insights about the traffic capacity of the different scenarios can be drawn. For both the four-way single-lane and the three-way single-lane intersections, \ac{Moveover} achieves the highest capacity thresholds, reaching total average densities of 0.48~\ac{veh/s} and 0.54~\ac{veh/s}, respectively, in the ideal case. These limits are maintained under a 5G architecture, while a negligible drop is observed with 4G. Specifically, in the first scenario, the proposed algorithm increases the capacity by 0.08~\ac{veh/s} compared to the most effective benchmark (the traffic light-based approach). Similarly, in the second scenario, \ac{Moveover} raises the saturation level by 0.15~\ac{veh/s} relative to the FIFO-based case, confirming its superior ability to delay the onset of congestion. 
The roundabout represents the only scenario where a benchmark, specifically the priority-based approach, reaches a higher saturation point (0.76~\ac{veh/s} on average) than \ac{Moveover}, which peaks at 0.72~\ac{veh/s} in the ideal case and 0.68 \ac{veh/s} under both 5G and 4G networks. The underlying reason for this difference in saturation points stems from \ac{Moveover}'s goal of preventing cars from coming to a complete stop. In contrast, the standard priority approach allows vehicles to come to a complete stop and queue at the yield line. While this stop-and-go behavior degrades travel times and emissions, it enables denser vehicle packing, allowing the system to process a marginally higher maximum volume of vehicles. Finally, in the four-way two-lane scenario, \ac{Moveover} achieves its highest overall capacity, reaching 1.24~\ac{veh/s} on average in the ideal case. Even when accounting for 5G network overhead, the saturation point remains high at 1.2~\ac{veh/s}, outperforming the traffic light-based approach, the most resilient benchmark, by 0.16~\ac{veh/s}. In stark contrast, the 4G architecture suffers a significant bottleneck, with its capacity dropping to 0.68~\ac{veh/s}. Because a two-lane configuration inherently handles a larger volume of vehicles, the amount of simultaneous data transmissions increases. Under 4G networks, the higher transmission delays create long processing queues at the server, eventually stalling the system. In the 5G scenario, the significantly lower latency ensures these server queues are cleared rapidly.

Figures~\ref{fig:messages_5G} and~\ref{fig:messages_4G} illustrate the empirical probability distribution of the number of messages required to complete the negotiation procedure across different scenarios, considering 5G and 4G connectivity, respectively. The reported values are obtained via \ac{SUMO} simulations, which are performed at the maximum sustainable density as discussed for Fig.~\ref{fig:density}.
Across both figures, the message exchange patterns remain consistent for all scenarios. The vast majority of negotiations are resolved with just two or four messages, while the need for six or eight messages is marginal.
Notably, in the four-way two-lane scenario, the use of 4G occasionally results in 10-message exchanges, which is a worst-case that does not occur when using 5G.

\subsection{Results over a real urban map with multiple intersections}\label{sec:multiintersection}

The effectiveness of \ac{Moveover} was also tested in the case of multiple intersections with a real layout. In particular, Fig.~\ref{fig:openstreetmapscenario} shows a real urban map with multiple intersections, derived from the suburban area of the Italian city of Verona using OpenStreetMap (OSM). This layout was simulated to assess the performance of our solution in a real-world scenario. The area features four intersections, specifically two four-way single-lane and two three-way single-lane junctions. The conflict zones are highlighted over the intersections; these are laid out in accordance with the models shown in Figs.~\ref{fig:model_a} and \ref{fig:model_b}, with their shapes adapted to the specific geometries of the junctions.
The scenario is simulated under two traffic management methods: the priority case, which is applied to all four intersections in the real urban map, and \ac{Moveover} with ideal communications. 

Results related to travel time and CO$_2$ emissions are depicted in Fig.~\ref{fig:multiNetwork}. These are obtained simulating the scenario with a total density of 0.5~\ac{veh/s}. As observable, our solution yields clear benefits compared to the benchmark across both metrics. In particular, looking at the average values it is observed a decrease of approximately 25\% of the travel time, which drops from 51.3~s to 37.7~s, and of the CO$_2$ emissions, which reduce from 0.16~kg to 0.12~kg.

\section{Conclusion}\label{sec:conclusions}

In this paper, we presented Moveover, a novel algorithm for autonomous intersection management utilizing V2N communications. The system effectively distributes the computational load by allowing each CAV to autonomously generate its own mobility profile while a local controller manages conflict zone scheduling. Extensive SUMO simulations validated the algorithm across diverse intersection models, including four-way, three-way, and roundabouts, as well as a real-world multi-intersection urban scenario. The evaluation yielded several key findings:

\begin{itemize}
    \item Moveover consistently outperformed traditional priority, traffic light, and FIFO scheduling methods, significantly reducing travel times and CO$_2$ emissions across nearly all tested densities.

    \item When applied to a simulated suburban map of Verona under a density of 0.5 vehicles per second, Moveover decreased average travel times and emissions by approximately 25\% compared to standard priority rules.

    \item Even when accounting for the latency introduced by 4G and 5G technologies, Moveover achieves robust performance even under heavy traffic volumes.
    
\end{itemize}

Future work will investigate how to adapt the proposed system to the presence of non-connected vehicles and vulnerable road users.






\appendix
\section{Design of the negotiation and conflict zones}\label{sec:zonedesign}
\begin{figure}[t]
\centering
\includegraphics[width=0.9\columnwidth]{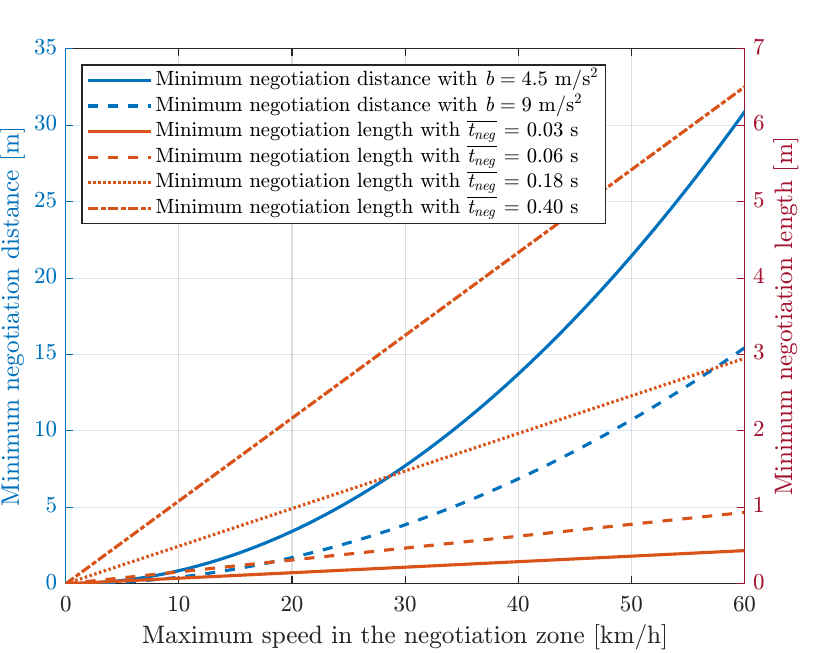}
\caption{Minimum negotiation distance and minimum negotiation length varying the maximum speed allowed in the negotiation zone.}
\label{fig:negotiation}
\end{figure}

 For the negotiation zones, the most critical factor is the distance between the end of the negotiation zone and the %
entrance of the intersection along the same lane, which we define as the \textit{minimum negotiation distance}. This distance must be sufficient for a vehicle to brake and come to a complete stop before reaching the intersection in case a mobility profile cannot be agreed upon. This requirement implies two possible design approaches: (i) specify a maximum allowable speed at the entrance of the negotiation zone and compute the corresponding minimum negotiation distance, or (ii) set the minimum negotiation distance in advance and derive the maximum allowable speed accordingly. The blue curves and left y-axis of Fig.~\ref{fig:negotiation} illustrates the relationship between the minimum negotiation distance and the maximum speed within the negotiation zone, assuming a constant braking deceleration $b$ of either 4.5~m/s$^2$ (standard) or 9~m/s$^2$ (emergency). With these assumptions, the minimum negotiation distance $\underline{d_\text{neg}}$ can be  calculated as a function of the maximum speed in the negotiation zone $\overline{v_\text{neg}}$ and the deceleration $b$ as \
\begin{equation}
     \underline{d_\text{neg}} ={\overline{{v_\text{neg}}}^2}/{2 b}\;.
\end{equation}
Given the values in Fig.~\ref{fig:negotiation}, 
if the maximum speed in the negotiation zone is for example 50~km/h, the minimum negotiation distance must be at least 22~m to allow comfortably braking in case of negotiation failure. 
As another example, if the road layout only permits a minimum negotiation distance of 10~m (e.g., due to proximity to a previous intersection), then the maximum speed in the negotiation zone must be limited to 34~km/h under standard deceleration.

A second design aspect concerns the distance between the beginning and the end of the negotiation zone, which we refer to as the \textit{negotiation length}. This length must be sufficient to ensure that the communication protocol, detailed in Section~\ref{sec:commprotocol}, is completed with sufficiently high probability (otherwise the backup mode is entered).  
Given the maximum speed of the vehicle inside the negotiation zone $\overline{v_\text{neg}}$ and the maximum time assumed for the negotiation  $\overline{t_\text{neg}}$, the minimum negotiation length $\underline{l_\text{neg}}$ can be expressed as
\begin{equation}
    \underline{l_\text{neg}} = \overline{v_\text{neg}} \cdot \overline{t_\text{neg}}\;.
\end{equation}
The red curves and left y-axis of Fig.~\ref{fig:negotiation} illustrate this relationship, showing the minimum negotiation length as a function of the maximum speed in the negotiation zone, assuming maximum negotiation duration of 0.03, 0.06, 0.18, and 0.4 s. These values refers to the four cases shown in Fig.~\ref{fig:sojourn_time}, assuming an average controller utilization of 50\%. For instance, for a maximum speed of 50 km/h, the negotiation zone must have a minimum length of 0.4, 0.8, 2.5, and 5.5 m for the various maximum negotiation durations.

Focusing on the conflict zones, their number and shape need to be optimized in order to balance two objectives: minimizing the risk of collisions, while maximizing the number of vehicles that can cross the intersection in the unit of time. The design depends on the specific intersection layout, including its type (e.g., four roads, three roads, roundabout), the number of lanes per road, the directions allowed, etc. Examples corresponding to the scenarios investigated in this work are provided in Fig.~\ref{fig:model} and discussed in more detail in Section~\ref{sec:simulation}.

\section{Discussion of the messages based on current ETSI standardization}\label{sec:ETSImessage}

The specific implementation of our protocol using standard messages requires further work. In this Appendix, we anyway use the information available to preliminary discuss this point and evaluate the size of the exchanged messages. For this purpose, we used a software called ASN.1 Studio~\cite{asn1studio} that allows to compute the length of a message based on its abstract syntax notation one (ASN.1) specification.

The specification of maneuver coordination messages is still under development in ETSI at the time of writing. Based on what is currently available, the MCM will include: 1) a standard C-ITS header and standard fields indicating the generation time and reference position (approximately 23 bytes); 2) more specific fields indicating the station type and station role (approximately 1 byte); and 3) a specific container for all additional information. The specific container in turn includes either a maneuver proposal or maneuver advice. 

When applied to our protocol from the CAVs to the controller, the messages may consist in a proposal, which includes: i) indications about the maneuver type, which is a proposal in an agreement seeking to use the ETSI terminology; ii) a description of the status of the vehicle; and iii) the description of the maneuver in terms of waypoints describing the trajectory; per each waypoint, at least the time instant, position and vehicle speed are indicated. If we consider 2 bytes for the maneuver type, 10 bytes for the vehicle status, 40 waypoints with 11 bytes per each waypoint, and add the 24 bytes detailed above, we obtain a message of less than 500 bytes.

When applied to our protocol from the controller to the CAVs, the messages may consist in a response with \ac{TRR} reservation, which includes: i) indications about the maneuver type, which is a response in an agreement seeking to use the ETSI terminology; ii) a description of the status of the vehicle; and iii) the definition of the TRRs to use; per each TRR, at least the type, size and interval of reservation are indicated. With approximately 2 bytes for the maneuver type, 10 bytes for the vehicle status, 10 TRRs with around 9 bytes per each TRR, and adding the 24 bytes detailed above, we obtain a message of less than 130 bytes.

If the negotiation fails for any reason, for instance a vehicle does not complete the negotiation procedure in time and the backup mode is triggered, a message requesting the cancellation of the maneuver can be sent by either the CAV or the controller. The message includes: i) indications about the maneuver type, which is a cancellation request to use the ETSI terminology; ii) a description of the status of the vehicle; and iii) the description of the maneuver, which in this case is empty. The first two fields combine for a total of around 12 bytes, while the third is only 5 bits. Adding the 24 bytes discussed at the start, the resulting message has a length of no more than 40 bytes.

\section*{Acknowledgment} This work was partially supported by the European Union under the Italian National Recovery and Resilience Plan (NRRP) of NextGenerationEU, partnership on ``Telecommunications of the Future'' (PE00000001 - program \\
\noindent``RESTART''), project MoVeOver and was partially funded by the European Union through the project CONNECT under grant agreement no. 101069688.

\bibliographystyle{elsarticle-num}

\bibliography{biblio}








\end{document}